\documentclass[fleqn,12pt]{article}
\usepackage{graphics,epsfig}

\newcommand{\be}[1]{\begin{equation} \label{(#1)}}
\newcommand{\ee}{\end{equation}}
\newcommand{\baq}[1]{\begin{eqnarray} \label{(#1)}}
\newcommand{\eaq}{\end{eqnarray}}

\newcommand{\rf}[1]{(\ref{(#1)})}

\def \noi {\noindent}

\def\lsim{\raise0.3ex\hbox{$\;<$\kern-0.75em\raise-1.1ex\hbox{$\sim\;$}}}
\def\gsim{\raise0.3ex\hbox{$\;>$\kern-0.75em\raise-1.1ex\hbox{$\sim\;$}}}

\def\beq   {\begin{equation}}
\def\eeq   {\end{equation}}
\def\beqd  {\begin{displaymath}}
\def\eeqd  {\end{displaymath}}
\def\beqaa {\begin{eqnarray}}
\def\eeqaa {\end{eqnarray}}

\def\ti  {\tilde}

\def\sg  {\ti g}

\def\nt  {\tilde\chi^0}
\def\ch  {\tilde\chi^\pm}

\def\a   {\alpha}
\def\b   {\beta}
\def\t   {\theta}

\def\gev {\rm ~GeV}

\def\sz{\ifmmode{\tilde{\chi}^0} \else{$\tilde{\chi}^0$} \fi}
\def\sw{\ifmmode{\tilde{\chi}} \else{$\tilde{\chi}$} \fi}

\newcommand{\AddrVienna}{
\it  Faculty of Physics, Universit\"at Wien, 
A-1090 Vienna, Austria \\}

\newcommand{\AddrAHEP}{
\it  AHEP Group, Instituto de Fisica Corpuscular - C.S.I.C., Universidad de Valencia, \\ 
Edificio Institutos de Investigacion, Apt. 22085, 
E-46071 Valencia, Spain \\}

\newcommand{\AddrCFTP}{
\it Departamento de F\'isica and CFTP, 
Instituto Superior T\'ecnico \\ Av. Rovisco Pais 1, 
1049-001 Lisboa, Portugal \\}

\newcommand{\AddrGAKUGEI}{%
 \it Department of Physics, Tokyo Gakugei University, Koganei, 
Tokyo 184-8501, Japan\\}

\newcommand{\AddrHEPHY}{%
 \it Institut f\"ur Hochenergiephysik der \"Osterreichischen Akademie 
der Wissenschaften, A-1050 Vienna, Austria\\}

\newcommand{\AddrWuerzburg}{%
 \it
Institut f\"ur Theoretische Physik and Astrophysik, Universit\"at W\"urzburg, 
D-97074 W\"urzburg, Germany\\}

\begin{document}

\begin{flushright}
%   TGU-35
% \\UWThPh-2005-13
% \\ CFTP/09-nnn
% \\HEPHY-PUB 811/05
% \\IFIC/05-31
% \\ZU-TH 10/05
\end{flushright}

\begin{center}  
  \textbf{\Large 
Impact of squark generation mixing on the search for gluinos at LHC}\\[10mm]

{A. Bartl${}^{1,2}$, K.~Hidaka${}^3$, K.~Hohenwarter-Sodek${}^1$, 
T. Kernreiter${}^4$, W.~Majerotto${}^5$ and W. Porod${}^6$
} 
\vspace{0.1cm}\\ 
$^1$ \AddrVienna
$^2$ \AddrAHEP
$^3$ \AddrGAKUGEI
$^4$ \AddrCFTP
$^5$ \AddrHEPHY
$^6$ \AddrWuerzburg
\end{center}

%\bigskip
\noindent

\begin{abstract}
\noindent
We study gluino decays in the Minimal Supersymmetric Standard Model (MSSM) with 
squark generation mixing. We show that the effect of this mixing on 
the gluino decay branching ratios can be very large in a significant part of the 
MSSM parameter space despite the very strong experimental constraints 
on quark flavour violation (QFV) from B meson observables. Especially we find 
that under favourable conditions the branching ratio of the the QFV gluino 
decay $\sg \to$ $c$ $\bar t$ ($\bar c$ $t$) $\nt_1$ can be as large as $\sim$ 50\%.
We also find that the squark generation mixing can result in a multiple-edge 
(3- or 4-edge) structure in the charm-top quark invariant mass distribution. 
The appearance of this remarkable structure provides an additional 
powerful test of supersymmetric QFV at LHC. These could have an important impact 
on the search for gluinos and the determination of the MSSM parameters at LHC.
\end{abstract}

\newpage 

%-----------------------------------------------------------------------------
\section{Introduction}
%-----------------------------------------------------------------------------

The search for supersymmetric (SUSY) particles will have  
a very high priority at the Large Hadron Collider (LHC) at CERN. 
If weak scale SUSY is realized in nature, gluinos 
and squarks, the  SUSY partners  of gluons  and quarks, will have high 
production rates for masses up to O(1 TeV).
The main decay modes of gluinos and squarks are 
usually assumed to be quark-flavour conserving (QFC). 
However, the squarks are not necessarily quark-flavour eigenstates 
and they are in general mixed by a $6 \times 6$ matrix. In this case 
quark-flavour violating (QFV) decays of gluinos and squarks could occur.
\\
The effect of QFV in the squark sector on reactions at colliders has been 
studied only in a few publications. 
The pair production of quarks with different flavours at the LHC is studied in 
\cite{QFV_quark_pair@LHC}. 
%\cite{Liu:2004bb,Bejar:2005kv,Eilam:2006rb,Cao:2007dk,LopezVal:2007rc,Bejar:2008ub}. 
%
The QFV effect can also be probed in the top quark decay 
\cite{QFV_top_decay}. 
%\cite{Li:1993mg,Couture:1994rr,Yang:1993rb,Lopez:1997xv,Guasch:1999jp,
%DiazCruz:2001gf,Cao:2006xb}. 
%
Moreover, QFV Higgs decays can have rates accessible at future colliders, 
see e.g. \cite{QFV_Higgs_decay}.
%see e.g. \cite{Curiel:2002pf,Demir:2003bv,Curiel:2003uk,Bejar:2004rz,Hahn:2005qi}.
%
In all of these studies the external particles of the reactions are Standard Model 
(SM) particles (or SUSY Higgs bosons).  This means that the effect of QFV 
in the squark sector is induced only by SUSY particle (sparticle) loops. \\
In sparticle reactions, on the other hand, the effect of QFV in the squark 
sector may be especially strong as they already occur at tree-level. 
The QFV decay $\tilde t_1\to c\tilde \chi_1^0$ \cite{QFV_decay_in_MFV} and 
QFV gluino decays \cite{Porod:2002wz} were 
%The QFV decay $\tilde t_1\to c\tilde \chi_1^0$ \cite{Hikasa:1987db,
%Han:2003qe,Lunghi:2006uf} and QFV gluino decays \cite{Porod:2002wz} were 
studied in the scenario of minimal flavour violation (MFV), where the  
only source of QFV is the mixing due to the Cabibbo-Kobayashi-Maskawa (CKM) matrix.
Note that the decay $\tilde t_1\to c\tilde \chi_1^0$ is actually 
the standard Tevatron search mode for light top-squarks. 
In \cite{Bozzi:2007me, QFV_squark_decay} squark pair production
%In \cite{Bozzi:2007me,Fuks:2008ab,Kribs:2009zy} squark pair production
and their decays at LHC have been analyzed including also the effect of 
the squark generation mixing.

In the present Letter, we study the effect of mixing between 
the second and third squark generations in its most general form. 
More precisely, we study the influence of the mixing of charm squark and top squark 
on the gluino and squark decays. In particular, we calculate the branching ratios of 
the following gluino decays into two quarks plus neutralino via up-type squark decay 
(see Fig.1) 
\footnote{
As we always sum over the particles and antiparticles of 
the (s)quarks, we do not indicate if it is a particle or its 
antiparticle: $q q'$ (with $q \neq q'$) means $q \bar{q'}$ and $\bar{q} q'$, 
and $q q$ means $q \bar{q}$, e.g. $B(\sg \to c t \nt_1) 
\equiv B(\sg \to c \bar t \nt_1) + B(\sg \to \bar c t \nt_1)$.
}:
\be{eq:decchain1}
\tilde g \to \tilde u_i~c \to c~t~ \tilde\chi^0_{1}~\quad
{\rm and}
\quad \tilde g \to \tilde u_i~t \to c~t~ \tilde\chi^0_{1}~.
\ee
%
%\baq{eq:react}                                                         <===
%\tilde g &\to& \tilde u_i c                                            <===
%             \to c t \tilde\chi^0_j\nonumber\\[2mm]                    <===
%\tilde g &\to& \tilde u_i t                                            <===
%             \to c t \tilde\chi^0_j.                                   <===
%\eaq                                                                   <===
%
We show that the QFV gluino decay branching ratio B($\sg \to c t \nt_1$) 
can be very large (up to $\sim$ 50\%) due to the squark generation mixing 
in a significant part of the MSSM parameter space despite the very strong 
experimental constraints from B factories, Tevatron and LEP 
\footnote{This is in analogy 
to the case of lepton flavour violating (LFV) sneutrino decays due to 
slepton generation mixing \cite{LFVsnu}.
}. 
%experimental constraints from B factories, Tevatron and LEP, in analogy 
%to the case of lepton flavour violating (LFV) sneutrino decays due to 
%slepton generation mixing \cite{LFVsnu}.
%}
%
%experimental constraints from B factories, Tevatron and LEP, where those from 
%$b \to s \gamma$ and $B^0_s-\bar B^0_s$ mixing are especially important 
%\cite{Hurth_Porod}. \\
%
We also study the effect of the squark generation mixing on the invariant mass 
distributions of the two quarks from the gluino decay at LHC. We show that 
it can result in novel multiple-edge structures in the distributions 
\footnote{
This is in analogy to the case of LFV neutralino decays due to slepton 
generation mixing \cite{LFV}.
}. \\
%in analogy to the case of LFV neutralino decays due to slepton 
%generation mixing \cite{LFV}. \\
%
These effects could have an important impact on the search for gluinos and the MSSM 
parameter determination at LHC.
%Here note that the establishment of squark flavour violation in the 
%soft-SUSY-breaking sector would be very important in our understanding of 
%supersymmetry breaking.

%-----------------------------------------------------------------------------
\section{Squark mixing with flavour violation}
%-----------------------------------------------------------------------------
%
%-----------------------------------------------------------------------------
%\section{Squark mixing with non--minimal flavor violation}
%-----------------------------------------------------------------------------

%First we summarize the MSSM parameters in our analysis.                        <===
Here we summarize the MSSM parameters in our analysis.
The most general up-type squark mass matrix including left-right mixing
as well as quark-flavour mixing in the super-CKM basis of 
$\tilde u_{0\gamma}=
(\tilde u_L,\tilde c_L,\tilde t_L,\tilde u_R,\tilde c_R,\tilde t_R)$, 
$\gamma=1,\dots,6$, is \cite{Allanach:2008qq}
%
% \bibitem{Allanach:2008qq}
%    B.~Allanach {\it et al.},
%    %``SUSY Les Houches Accord 2,''
%    Comput.\ Phys.\ Commun.\  {\bf 180} (2009) 8 
%    [arXiv:0801.0045 [hep-ph]].
%

\baq{eq:SquarkMassMatrix}
M^2_{\tilde u}=\left(\begin{array}{ccc}
M^2_{\tilde u LL} & (M^2_{\tilde u RL})^\dagger\\[5mm]
M^2_{\tilde u RL} & M^2_{\tilde u RR}
\end{array}\right)~,
\eaq
where the three $3\times3$ matrices read
\begin{eqnarray}
(M^2_{\tilde u LL})_{\alpha\beta} & = & 
M^2_{Q_u \alpha\beta}+\left[(\frac{1}{2}-\frac{2}{3}\sin^2\theta_W)
%(K\cdot M_Q^2\cdot K^\dagger)_{\alpha\beta}+\left[(\frac{1}{2}-\frac{2}{3}\sin^2\theta_W)
\cos2\beta~m_Z^2+m_{u_\alpha}^2\right]\delta_{\alpha\beta},
\label{eq:LL}\\[3mm]
(M^2_{\tilde u RR})_{\alpha\beta} & = & M_{U\alpha\beta}^2
+\left[\frac{2}{3}\sin^2\theta_W\cos2\beta~
m_Z^2+m_{u_\alpha}^2\right] \delta_{\alpha\beta}~,
\label{eq:RR}\\[3mm]
(M^2_{\tilde u RL})_{\alpha\beta} & = & (v_2/\sqrt{2} ) A_{U\beta\alpha}-
m_{u_\alpha} \mu^*\cot\beta~\delta_{\alpha\beta}~.
\label{eq:RL}
\end{eqnarray}
The indices $\alpha,\beta=1,2,3$ characterize the quark flavours $u,c,t$, respectively.
$M_{Q_u}^2$ and $M_U^2$ are the Hermitean soft-SUSY-breaking mass matrices for the left 
and right up-type squarks, respectively. Note that in the super-CKM basis one has 
$M_{Q_u}^2 = K\cdot M_Q^2\cdot K^\dagger $ due to the SU(2) symmetry, where $M_Q^2$ is 
the Hermitean soft-SUSY-breaking mass matrix for the left down-type squarks and 
$K$ is the CKM matrix. 
Note also that $M_{Q_u}^2 \simeq M_Q^2$ as $K \simeq 1$. 
%$K$ is the CKM matrix. Note also that $M_{Q_u}^2 \simeq M_Q^2$ as $K \simeq 1$. 
$A_U$ is the soft-SUSY-breaking trilinear coupling matrix of the up-type squarks: 
%$A_U$ is the trilinear soft SUSY breaking coupling matrix of the up-type squarks: 
${\mathcal L}_{\rm int}=
-(A_{U\alpha\beta} \tilde u^\dagger_{R\beta}\tilde  u_{L\alpha} H^0_2 + h.c.) + \cdots$.
$\mu$ is the higgsino mass parameter. 
$v_{1,2}$ are the vacuum expectation values of the Higgs fields with 
$v_{1,2}/\sqrt{2} \equiv \langle H^0_{1,2}\rangle$, and $\tan\beta \equiv v_2/v_1$. 
%$v_1$ and $v_2$ are the vacuum
%expectation values (VEVs) of the Higgs fields with $v_1 \equiv \langle H^0_1\rangle$,
%$v_2 \equiv \langle H^0_2\rangle$, and $\tan\beta \equiv v_2/v_1$ 
%
%\footnote{
%Note that the VEVs $v_{1,2}$ are defined as $v_{1,2}/\sqrt{2} \equiv \langle H^0_{1,2}\rangle$ 
%in \cite{Allanach:2008qq}.
%
%As for the definition of the VEVs $v_{1,2}$, $v_{1,2} \equiv \langle H^0_{1,2}\rangle$ in this 
%article while $v_{1,2}/\sqrt{2} \equiv \langle H^0_{1,2}\rangle$ in \cite{Allanach:2008qq}.
%}. 
%
$m_{u_\a}$ $(u_\a=u,c,t)$ are the physical quark masses.\\
The physical mass eigenstates $\tilde u_i$, $i=1,\dots,6$, are given
by $\tilde u_i=R^{\tilde u}_{i\alpha}\tilde u_{0\alpha}$.
{bf 
The mixing matrix $R^{\tilde u}$ and the mass eigenvalues 
are obtained by an unitary transformation 
%The $6 \times 6$ mixing matrix $R^{\tilde u}$ and the mass eigenstates            <===
%$\tilde u_i$ are obtained by an unitary transformation                            <===
$R^{\tilde u} M^2_{\tilde u} R^{\tilde u\dagger}=
{\rm diag}(m_{\tilde u_1},\dots,m_{\tilde u_6})$, where $m_{\tilde u_i}<m_{\tilde u_j}$
for $i<j$. \\
%Quark-flavour violation is induced by off--diagonal entries in the matrices $M^2_{Q_u}, <===
%M^2_U$ and $A_U$, i.e. squark generation mixing terms.                                  <===
%For instance, a non--zero $A_{U32}$ ($A_{U23}$)                                         <===
%gives rise to $\tilde c_R-\tilde t_L$ ($\tilde t_R-\tilde c_L$) mixing.                 <===
}
Having in mind that $M_{Q_u}^2 \simeq M_Q^2$, we define the QFV parameters 
$\delta^{uLL}_{\alpha\beta}$, $\delta^{uRR}_{\alpha\beta}$ 
and $\delta^{uRL}_{\alpha\beta}$ $(\alpha \neq \beta)$ as follows \cite{Gabbiani}: 
\begin{eqnarray}
\delta^{uLL}_{\alpha\beta} & \equiv & M^2_{Q \alpha\beta} / \sqrt{M^2_{Q \alpha\alpha} M^2_{Q \beta\beta}}~,
%
%(K\cdot M_Q^2\cdot K^\dagger)_{\alpha\beta} & \equiv & \sqrt{M^2_{Q \alpha\alpha} M^2_{Q \beta\beta}}~
%\delta^{uLL}_{\alpha\beta}~,
%
\label{eq:InsLL}\\[3mm]
\delta^{uRR}_{\alpha\beta} &\equiv& M^2_{U \alpha\beta} / \sqrt{M^2_{U \alpha\alpha} M^2_{U \beta\beta}}~,
\label{eq:InsRR}\\[3mm]
\delta^{uRL}_{\alpha\beta} &\equiv& (v_2/\sqrt{2} ) A_{U\beta\alpha} / \sqrt{M^2_{U \alpha\alpha} M^2_{Q \beta\beta}}~.
\label{eq:InsRL}
\end{eqnarray}
The down-type squark mass matrix can be analogously parametrized as the up-type 
squark mass matrix \cite{Allanach:2008qq}. 
As $M_Q^2 \simeq M_{Q_u}^2$, one has 
$(M^2_{\tilde d LL})_{\alpha\beta} \simeq (M^2_{\tilde u LL})_{\alpha\beta}$ 
for $\alpha \neq \beta$.
We do not introduce additional QFV terms in the down-type squark mass matrix.

%As a consequence, non--minimal FV in the left--left mixing block will have
%an important influence on observables of B--physics.
%This is in contrast to FV in the right--right and left--right mixing
%blocks whose parameters are far less restricted by 
%current measurements.
%We will be more detailed on this point in the next section
%where we give the constraints we impose on the model parameters.

The properties of the charginos $\ch_i$ ($i=1,2$, $m_{\ch_1}<m_{\ch_2}$) 
and neutralinos $\nt_k$ ($k=1,...,4$, $m_{\nt_1}< ...< m_{\nt_4}$)  
are determined by the parameters $M_2$, $M_1$, $\mu$ and $\tan\b$, 
where $M_2$ and $M_1$ are the SU(2) and U(1) gaugino masses, respectively. 
Assuming gaugino mass unification including the gluino mass $m_{\ti g}=M_3$, 
we take $M_1=(5/3)\tan^2\t_W M_2$. 
%Assuming gaugino mass unification we take $M_1=(5/3)\tan^2\t_W M_2$. 

%-----------------------------------------------------------------------------
\section{Constraints}\label{sec:constraints}
%-----------------------------------------------------------------------------

In our analysis, we impose the following conditions
on the MSSM parameter space in order to respect experimental
and theoretical constraints:

\renewcommand{\labelenumi}{(\roman{enumi})} 
% set counter to roman numbers
\begin{enumerate}
\item Constraints from the B-physics experiments relevant mainly for
      the mixing between the second and third generations of squarks
      \footnote{
      We do not consider the experimental constraints from $b \to s g$ and 
      $b \to s \nu \bar\nu$ since they have large uncertainties. We do not include 
      the constraints from the experimental data on $B(B_d \to \mu^+\mu^-)$, 
      $B(b \to d ~ l^+ l^-)$, $\Delta M_{B_d}$ and $\Delta M_{D^0}$ 
      as they practically do not constrain the 2nd and 3rd generation squark mixing 
      which we are interested in here.
      }: \\
      $3.03 \times 10^{-4} <  B(b \to s ~\gamma) < 4.01 \times 10^{-4}$ (95\% CL) 
                                                         \cite{Chang_ICHEP2008},
      $0.60 \times 10^{-6} <  B(b \to s ~ l^+l^-) < 2.60 \times 10^{-6}$ 
                                                  with $l=e ~{\rm or} ~\mu$ (95\% CL) 
                                                         \cite{b->sl+l-},
      $B(B_s \to \mu^+\mu^-) < 4.8 \times 10^{-8}$ (90\% CL) 
                                     \cite{Chang_ICHEP2008},
      $|R_{B\tau\nu}^{SUSY} - 1.77| < 1.27$ (95\% CL) 
      with $R_{B\tau\nu}^{SUSY} \equiv B^{SUSY}(B_u^- \to \tau^- {\bar\nu}_\tau) / 
      B^{SM}(B_u^- \to \tau^- {\bar\nu}_\tau) \simeq (1 - (\frac{m_{B^+}\tan\b}{m_{H^+}})^2)^2$ 
                                                         \cite{Hara_ICHEP2008}. 
      Moreover we impose the following condition on the SUSY prediction:
      $|\Delta M_{B_s}^{SUSY} - 17.77| < ((0.12 \times 1.96)^2 + 3.3^2)^{1/2} ~ps^{-1} 
      = 3.31 ~ps^{-1}$ (95\% CL),
      where we have combined the experimental error of $0.12  ps^{-1}$
      (at 68\% CL) \cite{DMBs_CDF} quadratically with the theoretical uncertainty
      of $3.3 ps^{-1}$ (at 95\% CL) \cite{DMBs_theoretical_error}. 
\item The experimental limit on SUSY contributions to the electroweak $\rho$ parameter \cite{rho_parameter}: 
$\Delta\rho(SUSY)<0.0012$.
\item The LEP limits on the SUSY particle masses 
        \cite{LEP}:
%        \cite{{LEP: ICHEP2002},{LEP: Higgs Working Group}}:
        $m_{\ch_1} > 103$ GeV, $m_{\nt_1} > 50$ GeV,
        $m_{\ti{u}_1,\ti{d}_1} > 100$ GeV, $m_{\ti{u}_1,\ti{d}_1} > m_{\nt_1}$,
        $m_{A^0} > 93~{\rm GeV}$, $m_{h^0}>110$~GeV, where $A^0$ is the CP-odd 
        Higgs boson and $h^0$ is the lighter CP-even Higgs boson. 
%
%For the lightest Higgs boson mass we require $m_{h^0}>111$~GeV,
%where we take into account theoretical uncertainties.
%
\item The Tevatron limit on the gluino mass \cite{m_gluino_ICHEP2008}:
        $m_{\tilde g}>$ 308 GeV.
\item The vacuum stability conditions for the trilinear coupling matrix \cite{Casas}:
\begin{eqnarray}
%|A_{U\alpha\alpha}|^2 &<&
%3~Y^2_{U\alpha}~(M^2_{Q\alpha\alpha}+M^2_{U\alpha\alpha}+m^2_2)~,
|A_{U\alpha\alpha}|^2 &<&
3~Y^2_{U\alpha}~(M^2_{Q_u \alpha\alpha}+M^2_{U\alpha\alpha}+m^2_2)~,
\label{eq:CCBfcU}\\[2mm]
|A_{D\alpha\alpha}|^2 &<&
3~Y^2_{D\alpha}~(M^2_{Q\alpha\alpha}+M^2_{D\alpha\alpha}+m^2_1)~,
\label{eq:CCBfcD}\\[2mm]
%|A_{U\alpha\beta}|^2 &<&
%Y^2_{U\gamma}~(M^2_{Q\alpha\alpha}+M^2_{U\beta\beta}+m^2_2)~, 
|A_{U\alpha\beta}|^2 &<&
Y^2_{U\gamma}~(M^2_{Q_u \alpha\alpha}+M^2_{U\beta\beta}+m^2_2)~, 
\label{eq:CCBfvU}\\[2mm]
|A_{D\alpha\beta}|^2 &<&
Y^2_{D\gamma}~(M^2_{Q\alpha\alpha}+M^2_{D\beta\beta}+m^2_1)~,
\label{eq:CCBfvD}
\end{eqnarray}
with 
$(\a\neq\b;\gamma={\rm Max}(\a,\b);\a,\b=1,2,3)$ and 
$m^2_1=(m^2_{H^\pm}+m^2_Z\sin^2\theta_W)\sin^2\beta-\frac{1}{2}m_Z^2$,
$m^2_2=(m^2_{H^\pm}+m^2_Z\sin^2\theta_W)\cos^2\beta-\frac{1}{2}m_Z^2$.
The Yukawa couplings of the up-type and down-type quarks are 
$Y_{U\alpha}=\sqrt{2}m_{u_\alpha}/v_2=\frac{g}{\sqrt{2}}\frac{m_{u_\alpha}}{m_W \sin\beta}$ 
$(u_\a=u,c,t)$ and 
$Y_{D\alpha}=\sqrt{2}m_{d_\alpha}/v_1=\frac{g}{\sqrt{2}}\frac{m_{d_\alpha}}{m_W \cos\beta}$ 
$(d_\a=d,s,b)$, 
with $m_{u_\a}$ and $m_{d_\a}$ being the running quark masses at the scale of $m_Z$ and 
$g$ the SU(2) gauge coupling. All soft-SUSY-breaking parameters are assumed to be given 
at the scale of $m_Z$. As SM input we take $m_W=80.4 \gev$, $m_Z=91.2 \gev$ and 
the on-shell top-quark mass $m_t=174.3 \gev$. 
We have found that our results shown in the following are fairly insensitive to $m_t$.
\end{enumerate}
We calculate the observables in (i)-(iii) by using the public code SPheno v3.0 
\cite{SPheno_B-physics_refs}.
%We calculate the observables in (i) and (ii) by using the formulae of Ref.~\cite{B-physics}.
Condition (i) except for $B(B^+_u\to \tau^+ \nu)$ strongly constrains the 2nd and 3rd 
%generation squark mixing parameters $M^2_{Q 23}, M^2_{U 23}, M^2_{D 23}, A_{U 23}, 
%A_{U 32}, A_{D 23}$ and $A_{D32}$; the constraints from $B(b \to s \gamma)$ and 
generation squark mixing parameters $M^2_{Q 23}, M^2_{D 23}, A_{U 23}, 
A_{D 23}$ and $A_{D32}$; the constraints from $B(b \to s \gamma)$ and 
$\Delta M_{B_s}$ are especially important \cite{Hurth_Porod}.
%
%Condition (i) except for $B(B^+_u\to \tau^+ \nu)$ strongly constrains the 2nd and 3rd 
%generation squark mixing parameters $M^2_{Q 23}, M^2_{U 23}$, $A_{U 23}$ and $A_{U 32}$.
%

%-----------------------------------------------------------------------------
\section{Quark flavour violating gluino decays}
%-----------------------------------------------------------------------------

We study the effect of the 2nd and 3rd generation squark mixing on the gluino decays. 
We focus on the QFV gluino decays of Eq.\rf{eq:decchain1} 
leading to the same final state $ c~t~ \tilde\chi^0_{1}$.
%
%We focus on the QFV gluino decays                                   <===
%\be{eq:decchain1}                                                   <===
%\tilde g \to \tilde u_i~c \to c~t~ \tilde\chi^0_{1}~\quad           <===
%{\rm and}                                                           <===
%\quad \tilde g \to \tilde u_i~t \to c~t~ \tilde\chi^0_{1}~,         <===
%\ee                                                                 <===
%%                                                                   <===
%leading to the same final state $ c~t~ \tilde\chi^0_{1}$.           <===
%
%\be{eq:FVdecay}
%\tilde g \to c~t~ \tilde\chi^0_{1}~.
%\ee
% 
We calculate the gluino and squark decay widths taking into account 
the following two--body decays: 
%For the calculation of the gluino and squark decay widths we
%take into account the following two--body decays 
%
\baq{eq:decaymodes}
\tilde g &\to& \tilde u_i~u_k,~\tilde d_i~d_k,\nonumber\\[2mm]
\tilde u_i &\to& u_k ~\tilde\chi^0_n, ~d_k ~\tilde\chi^+_m,~\tilde{d}_j~W^+,~
\tilde{u}_j~Z^0,~\tilde{u}_j~h^0,
\eaq
where $u_k=(u,c,t)$ and $d_k=(d,s,b)$.
The squark decays into the heavier Higgs bosons are kinematically 
forbidden in our scenarios studied below. 
The formulae for the two--body decays in \rf{eq:decaymodes} can be found in 
\cite{Bozzi:2007me}, except for the squark decays into the Higgs bosons for which 
we take the formulae of \cite{Bartl:2003pd} modified appropriately with the 
squark mixing matrix in the general QFV case. 

\begin{table}[t]
\begin{center}

\begin{tabular}{|c||c|c|c|} \hline
 $M^2_{Q\alpha\beta}$
& \multicolumn{1}{c|}{\scriptsize{${\beta=1}$}} 
& \multicolumn{1}{c|}{\scriptsize{$\beta=2$}} 
& \multicolumn{1}{c|}{\scriptsize{$\beta=3$}} \\\hline\hline
 \scriptsize{$\alpha=1$}
& \multicolumn{1}{c|}{$(920)^2$} 
& \multicolumn{1}{c|}{0} 
& \multicolumn{1}{c|}{0} \\\hline

 \scriptsize{$\alpha=2$}
& \multicolumn{1}{c|}{0} 
& \multicolumn{1}{c|}{$(880)^2$} 
& \multicolumn{1}{c|}{$(224)^2$} \\\hline

 \scriptsize{$\alpha=3$}
& \multicolumn{1}{c|}{0} 
& \multicolumn{1}{c|}{$(224)^2$} 
& \multicolumn{1}{c|}{$(840)^2$} \\\hline
\end{tabular}
\begin{tabular}{|c|c|c|c|c|c|} \hline
 
  \multicolumn{1}{|c|}{$M_1$} 
& \multicolumn{1}{c|}{$M_2$} 
& \multicolumn{1}{c|}{$m_{\sg}$} 
& \multicolumn{1}{c|}{$\mu$} 
& \multicolumn{1}{c|}{$\tan\beta$} 
& \multicolumn{1}{c|}{$m_{A^0}$} \\\hline\hline
 
  \multicolumn{1}{|c|}{139} 
& \multicolumn{1}{c|}{264} 
& \multicolumn{1}{c|}{800} 
& \multicolumn{1}{c|}{1000} 
& \multicolumn{1}{c|}{10} 
& \multicolumn{1}{c|}{800} \\\hline

\end{tabular}
\vskip0.2cm
\begin{tabular}{|c||c|c|c|} \hline
 $M^2_{D\alpha\beta}$
& \multicolumn{1}{c|}{\scriptsize{${\beta=1}$}} 
& \multicolumn{1}{c|}{\scriptsize{$\beta=2$}} 
& \multicolumn{1}{c|}{\scriptsize{$\beta=3$}} \\\hline\hline
 \scriptsize{$\alpha=1$}
& \multicolumn{1}{c|}{$(830)^2$} 
& \multicolumn{1}{c|}{0} 
& \multicolumn{1}{c|}{0} \\\hline

 \scriptsize{$\alpha=2$}
& \multicolumn{1}{c|}{0} 
& \multicolumn{1}{c|}{$(820)^2$} 
& \multicolumn{1}{c|}{0} \\\hline

 \scriptsize{$\alpha=3$}
& \multicolumn{1}{c|}{0} 
& \multicolumn{1}{c|}{0} 
& \multicolumn{1}{c|}{$(810)^2$} \\\hline
\end{tabular}
\hspace{0.6cm}
\begin{tabular}{|c||c|c|c|} \hline
 $M^2_{U\alpha\beta}$
& \multicolumn{1}{c|}{\scriptsize{${\beta=1}$}} 
& \multicolumn{1}{c|}{\scriptsize{$\beta=2$}} 
& \multicolumn{1}{c|}{\scriptsize{$\beta=3$}} \\\hline\hline
 \scriptsize{$\alpha=1$}
& \multicolumn{1}{c|}{$(820)^2$} 
& \multicolumn{1}{c|}{0} 
& \multicolumn{1}{c|}{0} \\\hline

 \scriptsize{$\alpha=2$}
& \multicolumn{1}{c|}{0} 
& \multicolumn{1}{c|}{$(600)^2$} 
& \multicolumn{1}{c|}{$(224)^2$} \\\hline

 \scriptsize{$\alpha=3$}
& \multicolumn{1}{c|}{0} 
& \multicolumn{1}{c|}{$(224)^2$} 
& \multicolumn{1}{c|}{$(580)^2$} \\\hline
\end{tabular}
%\\[0.5ex]
\vskip0.4cm
\caption{\label{tab1}
The MSSM parameters in our reference scenario with QFV.
All of $A_{U \a\b}$ and $A_{D \a\b}$ are set to zero.
All mass parameters are given in GeV.}
%All of the remaining parameters of the squark system
%are set to zero.}
\end{center}
\end{table}

\begin{table}[t]
\begin{center}

\begin{tabular}{|c|c|c|c|c|c|} \hline
 
  \multicolumn{1}{|c|}{$\tilde u_1$} 
& \multicolumn{1}{c|}{$\tilde u_2$} 
& \multicolumn{1}{c|}{$\tilde u_3$} 
& \multicolumn{1}{c|}{$\tilde u_4$} 
& \multicolumn{1}{c|}{$\tilde u_5$} 
& \multicolumn{1}{c|}{$\tilde u_6$} \\\hline\hline
 
  \multicolumn{1}{|c|}{558} 
& \multicolumn{1}{c|}{642} 
& \multicolumn{1}{c|}{819} 
& \multicolumn{1}{c|}{837} 
& \multicolumn{1}{c|}{897} 
& \multicolumn{1}{c|}{918} \\\hline

\end{tabular}
\begin{tabular}{|c|c|c|c|c|c|} \hline
 
  \multicolumn{1}{|c|}{$\tilde d_1$} 
& \multicolumn{1}{c|}{$\tilde d_2$} 
& \multicolumn{1}{c|}{$\tilde d_3$} 
& \multicolumn{1}{c|}{$\tilde d_4$} 
& \multicolumn{1}{c|}{$\tilde d_5$} 
& \multicolumn{1}{c|}{$\tilde d_6$} \\\hline\hline
 
  \multicolumn{1}{|c|}{800} 
& \multicolumn{1}{c|}{820} 
& \multicolumn{1}{c|}{830} 
& \multicolumn{1}{c|}{835} 
& \multicolumn{1}{c|}{897} 
& \multicolumn{1}{c|}{922} \\\hline

\end{tabular}
\vskip0.2cm
\begin{tabular}{|c|c|c|c||c|c|} \hline
 
  \multicolumn{1}{|c|}{$\tilde \chi^0_1$} 
& \multicolumn{1}{c|}{$\tilde \chi^0_2$} 
& \multicolumn{1}{c|}{$\tilde \chi^0_3$} 
& \multicolumn{1}{c||}{$\tilde \chi^0_4$} 
& \multicolumn{1}{c|}{$\tilde \chi^\pm_1$} 
& \multicolumn{1}{c|}{$\tilde \chi^\pm_2$} \\\hline\hline
 
  \multicolumn{1}{|c|}{138} 
& \multicolumn{1}{c|}{261} 
& \multicolumn{1}{c|}{1003} 
& \multicolumn{1}{c||}{1007} 
& \multicolumn{1}{c|}{261} 
& \multicolumn{1}{c|}{1007} \\\hline

\end{tabular}
\vskip0.4cm
\caption{\label{tab2}
Sparticles and corresponding masses (in GeV) in the scenario of 
Table \ref{tab1}.
}
\end{center}
\end{table}

We take $\tan\b, m_{A^0}, M_1, M_2, m_{\ti g}, \mu, M^2_{Q\a\b}, M^2_{U \a\b}, 
M^2_{D \a\b}, A_{U \a\b}$ and $A_{D \a\b}$ as the basic MSSM parameters at the 
weak scale. We assume them to be real. 
The QFV parameters are the squark generation 
mixing terms $M^2_{Q\a\b}$, $M^2_{U \a\b}$, $M^2_{D \a\b}$, $A_{U \a\b}$ and 
$A_{D \a\b}$ with $\a \neq \b$. 
%Note that the so-called minimal flavour violation (MFV) 
%corresponds to the case where all of these squark generation mixing terms are zero and 
%the CKM mixing matrix is the only source of flavour violation (QFV). 
As a reference scenario, we take the scenario given in Table \ref{tab1}.
This scenario is within the reach of LHC and satisfies the conditions (i)-(v).
For the observables in (i) and (ii) we obtain $B(b \to s \gamma)=3.57\times10^{-4}, 
~B(b \to s l^+l^-)= 1.59\times10^{-6}, ~B(b \to s \nu \bar\nu)=4.07\times10^{-5}, 
~B(B_s \to \mu^+\mu^-)=4.72\times10^{-9}, ~B(B^+_u \to \tau^+ \nu)=7.85\times10^{-5}, 
~\Delta M_{B_s}= 17.38 ~ps^{-1}$ and $\Delta\rho(SUSY)= 1.50\times10^{-4}$. 
%The resulting chargino and neutralino 
%masses are given by $m_{\ch_1}= $~GeV, $m_{\ch_2}= $~GeV and $m_{\nt_1}= $~GeV, 
%$m_{\nt_2}= $~GeV, $m_{\nt_3}= $~GeV, $m_{\nt_4}= $~GeV, respectively. 
The resulting masses of squarks, neutralinos and charginos  are given in Table \ref{tab2}. 
We show the up-type squark compositions in the flavour eigenstates in Table \ref{tab3}.\\
%The resulting masses of squarks, neutralinos and charginos  are given in Table \ref{tab2}. 
%We show the squark compositions in the flavour eigenstates in Table \ref{tab3}.\\
%
\begin{table}[t]
\begin{center}
%\hspace{2mm}   
\begin{tabular}{|c||c|c|c|c|c|c|}
		 \hline
			$R^{\tilde u}_{i\a}$
		  & $\ti u_L$ & $\ti c_L$ & $\ti t_L$ & $\ti u_R$ & $\ti c_R$ & $\ti t_R$ \\
		 \hline\hline
		  $\ti u_1$  & -0.001 & 0.005 & -0.029 & 0 & 0.728 & -0.685 \\
		  $\ti u_2$  & -0.002 & 0.008 & -0.040 & 0 & -0.686 & -0.727 \\
		  $\ti u_3$  & 0 & 0 & 0 & 1.0 & 0 & 0 \\
		  $\ti u_4$  & 0.128 & -0.583 & 0.801 & 0 & -0.007 & -0.045 \\
		  $\ti u_5$  & -0.181 & 0.782 & 0.597 & 0 & -0.003 & -0.021 \\
		  $\ti u_6$  & -0.975 & -0.221 & -0.005 & 0 & 0 & 0 \\
%		  $\ti u_1$  & -0.0012 & 0.0052 & -0.0286 & 0 & 0.7278 & -0.6852 \\
%		  $\ti u_2$  & -0.0018 & 0.0081 & -0.0401 & 0 & -0.6857 & -0.7267 \\
%		  $\ti u_3$  & 0 & 0 & 0 & 1.0 & 0 & 0 \\
%		  $\ti u_4$  & 0.1279 & -0.5832 & 0.8009 & 0 & -0.0065 & -0.0449 \\
%		  $\ti u_5$  & -0.1806 & 0.7816 & 0.5967 & 0 & -0.0025 & -0.0214 \\
%		  $\ti u_6$  & -0.9752 & -0.2213 & -0.0053 & 0 & 0 & 0.0003 \\

		 \hline
		 \end{tabular} 
		 
\vspace{3mm} 
%
%\begin{tabular}{|c||c|c|c|c|c|c|}
%		 \hline
%			$R^{\tilde d}_{i\a}$
%		  & $\ti d_L$ & $\ti s_L$ & $\ti b_L$ & $\ti d_R$ & $\ti s_R$ & $\ti b_R$ \\
%		 \hline\hline
%		  $\ti d_1$  & 0 & 0.1768 & -0.4860 & 0 & 0.0037 & -0.8559 \\
%		  $\ti d_2$  & 0 & 0.0091 & -0.0052 & 0 & 0.9999 & 0.0092 \\
%		  $\ti d_3$  & 0.0003 & 0 & 0 & 1.0 & 0 & 0 \\
%		  $\ti d_4$  & 0 & -0.4559 & 0.7302 & 0 & 0.0126 & -0.5088 \\
%		  $\ti d_5$  & 0 & -0.8722 & -0.4803 & 0 & 0.0045 & 0.0925 \\
%		  $\ti d_6$  & -1.0 & 0 & 0 & 0.0003 & 0 & 0 \\
%
%		 \hline
%		 \end{tabular} 

	\caption{\label{tab3} The up-type squark compositions in the flavour eigenstates, 
i.e. the mixing matrix $R^{\ti u}_{i\a}$ for the scenario of Table \ref{tab1}.}
%	\caption{\label{tab3} The squark compositions in the flavour eigenstates, 
%i.e. the mixing matrices $R^{\ti u}_{i\a}$ and $R^{\ti d}_{i\a}$ 
%for the scenario of Table \ref{tab1}.}

\end{center}
\end{table}
For the important branching ratios of the gluino and squark two-body decays we get 
$B(\sg \to \ti{u_1} c)= 0.481, ~B(\sg \to \ti{u_1} t)= 0.300, ~B(\sg \to \ti{u_2} c)= 0.207, 
~B(\sg \to \ti{u_2} t)= 0.0, ~{\rm and} ~B(\ti{u_1} \to c \nt_1)= 0.576, ~
B(\ti{u_1} \to t \nt_1)= 0.401, ~B(\ti{u_2} \to c \nt_1)= 0.495, ~B(\ti{u_2} \to t \nt_1)= 0.469$. 
This leads to the following gluino decay branching ratios: 
\begin{eqnarray}
   \hspace{-1.0cm} 
   B(\sg \to c t \nt_1) & \hspace{-0.2cm} = & \hspace{-0.4cm} \sum_{i=1,2} \hspace{-0.1cm} 
   \left[B(\sg \to \ti{u_i} c) B(\ti{u_i} \to t \nt_1) + 
   B(\sg \to \ti{u_i} t) B(\ti{u_i} \to c \nt_1)\right] \hspace{-0.1cm} = 0.463,
   \label{eq:Bgluino_ctN1}\\ %[3mm] 
   \hspace{-1.0cm}
   B(\sg \to c c \nt_1) & \hspace{-0.2cm} = & \hspace{-0.4cm} \sum_{i=1,2} \hspace{-0.1cm} 
   \left[B(\sg \to \ti{u_i} c) B(\ti{u_i} \to c \nt_1) \right] \hspace{-0.1cm} = 0.380,
   \label{eq:Bgluino_ccN1}\\ %[3mm] 
   \hspace{-1.0cm}
   B(\sg \to t t \nt_1) & \hspace{-0.2cm} = & \hspace{-0.4cm} \sum_{i=1,2} \hspace{-0.1cm} 
   \left[B(\sg \to \ti{u_i} t) B(\ti{u_i} \to t \nt_1) \right] \hspace{-0.1cm} = 0.120.
   \label{eq:Bgluino_ttN1} 
\end{eqnarray}
%
%
%This leads to a very large QFV gluino decay branching ratio 
%\beq \label{eq:Bgluino_ctN1}
%\hspace{-1.0cm} B(\sg \to c t \nt_1)= \hspace{-0.2cm}\sum_{i=1,2} \hspace{-0.1cm} 
%\left[B(\sg \to \ti{u_i} c) B(\ti{u_i} \to t \nt_1) + 
%B(\sg \to \ti{u_i} t) B(\ti{u_i} \to c \nt_1)\right] \hspace{-0.1cm} = 0.463. 
%\eeq
%
Note that the QFV gluino decay branching ratio of Eq. (\ref{eq:Bgluino_ctN1}) is very large.
The reason of this very large QFV gluino decay branching ratio is as follows: 
The gluino decays into squarks other than $\ti u_{1,2}$ are kinematically forbidden, 
and $\ti u_1$ , $\ti u_2$ are strong mixtures of the flavour eigenstates $\ti c_R$ 
and $\ti t_R$ due to the large $\ti c_R$ - $\ti t_R$ mixing term $M^2_{U 23} 
(= (224 \gev)^2)$ in this scenario. This results in the large branching ratios of 
$B(\sg \to \ti{u_i} c), B(\sg \to \ti{u_i} t) ~{\rm and} ~B(\ti{u_i} \to c \nt_1), 
B(\ti{u_i} \to t \nt_1)$ with $i=1,2$, except for the branching ratio of the decay 
$\sg \to \ti{u_2} t$ which is kinematically forbidden. Note that $\ti u_{1,2} 
(\sim \ti c_R + \ti t_R)$ couple to $\nt_1 (\simeq \ti B^0)$ and practically do not couple 
to $\nt_2 (\simeq \ti W^0)$, $\ch_1 (\simeq \ti W^\pm)$, and that $\nt_{3,4}$, $\ch_2$ are 
very heavy in this scenario. Here $\ti B^0$ and $\ti W^{0, \pm}$ are the U(1) 
and SU(2) gauginos, respectively.

We now study the basic MSSM parameter dependences of the QFV gluino and squark decay 
branching ratios for the reference scenario of Table \ref{tab1}.
In Fig.2 we show contours of $B(\sg \to c t \nt_1)$ in the ($\Delta M^2_U, M^2_{U 23}$) 
plane with $\Delta M^2_U \equiv M^2_{U 22} - M^2_{U 33}$. 
All basic parameters other than $M^2_{U 22}$ and $M^2_{U 23}$ are fixed as in our 
reference scenario defined in Table \ref{tab1}. 
%All basic parameters other than $M^2_{U 22}$ and $M^2_{U 23}, 
%M^2_{U 32}$ are fixed as in our reference scenario defined in Table \ref{tab1}. 
%Our reference scenario corresponds to $(\Delta M^2_U, M^2_{U 23}) = 
%(2.36 \times 10^4, 5 \times 10^4) \gev^2$. 
We see that the QFV decay branching ratio $B(\sg \to c t \nt_1)$ quickly increases 
up to $\sim$ 50\% with increase of the effective $\ti c_R - \ti t_R$ mixing angle 
$\tan(2\theta^{eff}_{23}) \equiv 2M^2_{U 23}/\Delta M^2_U$. 
%with 
%$\Delta M^2_U = m^2_{\ti c_R}-m^2_{\ti t_R} + m_t^2 - m_c^2 
%\sim m^2_{\ti c_R}-m^2_{\ti t_R}$. 
%
%$\Delta M^2_U \sim m^2_{\ti c_R}-m^2_{\ti t_R}$. 
%
%Note : DM^2_U = M^2_U22 - M^2_U33 = 600^2 - 580^2 = 23600
%
%       m(~c_R)^2 - m(~t_R)^2 = (600^2 -1250 + 1.3^2) - (580^2 -1250 + 174^2) 
%                             = 600^2 - 580^2 + 1.3^2 -174^2  = -6015
%
%       ---> DM^2_U = M^2_U22 - M^2_U33 =/= m(~c_R)^2 - m(~t_R)^2 !
%

In Fig.3 we present contours of $B(\sg \to c t \nt_1)$ in the $\delta^{uLL}_{23} - 
\delta^{uRR}_{23}$ plane where all of the conditions (i)-(v) except the 
$b \to s \gamma$ constraint are satisfied. 
For $b \to s \gamma$ we also show the corresponding branching ratio contours.
All basic parameters other than $M^2_{Q 23}$ and $M^2_{U 23}$ 
%All basic parameters other than $M^2_{Q 23}, M^2_{Q 32}$ and $M^2_{U 23}, M^2_{U 32}$ 
are fixed as in our reference scenario defined in Table \ref{tab1}. 
%Our reference scenario corresponds to $(\delta^{uLL}_{23}, \delta^{uRR}_{23}) = 
%(0.068, 0.144)$. 
We see that the QFV decay branching ratio 
$B(\sg \to c t \nt_1)$ increases quickly with increase of the $\ti c_R - \ti t_R$ mixing 
parameter $|\delta^{uRR}_{23}|$ and can be very large (up to $\sim$ 50 \%) in a significant 
part of the $\delta^{uLL}_{23} - \delta^{uRR}_{23}$ plane allowed by all of the 
conditions (i)-(v) including the $b \to s \gamma$ constraint. $B(\sg \to c t \nt_1)$ is 
insensitive to the $\ti c_L - \ti t_L$ mixing parameter $\delta^{uLL}_{23}$ and 
can be quite large ($\sim$ 50\%) in a sizable allowed range $0.03 \lsim \delta^{uLL}_{23} 
\lsim 0.12$.

Studying the branching ratios of the gluino and up-type squark two-body decays separately 
allows for a better understanding of their contributions to the QFV gluino decay 
$\tilde g \to c t\tilde\chi_1^0$. 
In Fig.4 we show the $\delta^{uRR}_{23}$ (i.e. ${\ti c}_R - {\ti t}_R$ mixing parameter) 
dependences of the gluino and squark decay branching ratios, where all basic parameters 
other than $M^2_{U 23}$ are fixed as in the scenario of Table \ref{tab1}. 
We see that $B(\sg \to c t \nt_1)$ increases quickly 
with increase of $|\delta^{uRR}_{23}|$ for $|\delta^{uRR}_{23}| \lsim 0.1$ and can be 
very large ($\sim$ 50\%) in a wide range of $\delta^{uRR}_{23}$.
This behaviour can be explained by an argument similar to that below 
Eq.(\ref{eq:Bgluino_ttN1}).
In Fig.4(b) [(c)] we see that $B(\sg \to \ti{u_i} c)$ and $B(\sg \to \ti{u_i} t)$ 
[$B(\ti{u_i} \to c \nt_1)$ and $B(\ti{u_i} \to t \nt_1)$] with $i=1,2$ are 
large in a wide range of $\delta^{uRR}_{23}$, except for $B(\sg \to \ti{u_2} t)$ 
which is kinematically suppressed. This leads to the very large $B(\sg \to c t \nt_1)$ 
in a wide range of $\delta^{uRR}_{23}$ (see Eq.(\ref{eq:Bgluino_ctN1})).
%Note that $m_{\ti u_1}$ ($m_{\ti u_2}$) decreases (increases) with the increase of 
%$|\delta^{uRR}_{23}|$, which explains the behaviour of the two-body decay branching 
%ratios for $|\delta^{uRR}_{23}| \gsim 0.1$. 
%Notice also that $\ti u_1=\ti c_R$ and $\ti u_2 \simeq \ti t_R$ for 
%$\delta^{uRR}_{23}=0$, which explains the behaviour around $\delta^{uRR}_{23}=0$. 

In Fig.5 we show the $\delta^{uRL}_{23}$ (i.e. ${\ti c}_R - {\ti t}_L$ mixing parameter) 
dependences of the gluino decay branching ratios, where all basic parameters other than 
$A_{U 32}$ are fixed as in the scenario of Table \ref{tab1}. 
We see that the QFV decay branching ratio $B(\sg \to c t \nt_1)$ can be quite large 
($\sim$ 30-50\%) in a wide range of $\delta^{uRL}_{23}$. $B(\sg \to c t \nt_1)$ decreases 
(down to $\sim 30\%$) and the quark-generation violating (QGV) decay branching ratio 
$B(\sg \to c b \ch_1)$ increases (up to $\sim 20\%$) with increase of $|\delta^{uRL}_{23}|$. 
%This behaviour can be explained as follows: 
%In this scenario the $\ch_1$ ($\simeq {\ti W}^\pm$ (wino)) couples to $\ti q_L$ and 
%its coupling to $\ti q_R$ is suppressed. 
%
%On the other hand, $\nt_1$ ($\simeq {\ti B}^0$ (bino)) couples much more strongly to 
%$\ti c_R$ and $\ti t_R$ than to $\ti c_L$ and $\ti t_L$. 
%
%On the other hand, $\nt_1$ ($\simeq {\ti B}^0$ (bino)) couples to $\ti c_R$ and $\ti t_R$ 
%and does not significantly couple to $\ti c_L$ and $\ti t_L$. 
%
Sizable $\delta^{uRL}_{23}$ (i.e. ${\ti c}_R - {\ti t}_L$ mixing parameter) 
induces  a sizable ${\ti t}_L$ component in ${\ti u}_{1,2} (\sim {\ti c}_R + {\ti t}_R)$, 
which enhances the widths $\Gamma({\ti u}_{1,2} \to b ~\ch_1(\simeq {\ti W}^\pm))$ and leads 
to a suppression of $B({\ti u}_{1,2} \to c ~\nt_1(\simeq {\ti B}^0))$ and 
$B({\ti u}_{1,2} \to t \nt_1)$. As a result 
$B(\sg \to c b \ch_1) = \sum_{i=1,2} B(\sg \to \ti{u}_i c) B(\ti{u}_i \to b \ch_1)$ 
\footnote{
Note that gluino decays into a down-type squark, such as $B(\sg \to \ti{d_i} b)$, are 
kinematically forbidden in this scenario and hence that such decays cannot contribute 
to $B(\sg \to c b \ch_1)$.
% \sum_{i=1,2} B(\sg \to \ti{d_i} b) B(\ti{d_i} \to c \ch_1) = 0
}
is enhanced for sizable $\delta^{uRL}_{23}$ while
$B(\sg \to c t \nt_1) = \sum_{i=1,2} 
   \left[B(\sg \to \ti{u}_i c) B(\ti{u}_i \to t \nt_1) + 
   B(\sg \to \ti{u}_i t) B(\ti{u}_i \to c \nt_1)\right]$ is suppressed.

As for the $\delta^{uRL}_{32}$ (i.e. ${\ti c}_L - {\ti t}_R$ mixing parameter) dependence 
of the gluino decay branching ratios, 
%where all basic parameters other 
%than $A_{U 23}$ are fixed as in the scenario of Table \ref{tab1}, 
%we have obtained similar results (including the allowed range) to those 
we have obtained similar results to those 
for the $\delta^{uRL}_{23}$ dependence in Fig.5. 
We have found that $B(\sg \to c t \nt_1)$ can be quite large ($\sim$ 30-50\%) in a 
wide allowed range $|\delta^{uRL}_{32}| \lsim 0.3$. 
$B(\sg \to c t \nt_1)$ decreases (down to $\sim 30\%$) and the QGV decay branching 
ratio $B(\sg \to s t \ch_1)$ increases (up to $\sim 5\%$) with the increase of 
$|\delta^{uRL}_{32}|$ while $B(\sg \to c b \ch_1)$ is small. 
%This behaviour can be explained by an argument similar to that in the case of 
%the $\delta^{uRL}_{23}$ dependence. 

%As for the $\delta^{uLL}_{23}$ (i.e. $\ti c_L - \ti t_L$ mixing parameter) 
%dependence plot of the gluino decay branching ratios, where all basic parameters 
%other than $M^2_{Q 23}$ are fixed as in the scenario of Table \ref{tab1}, we have 
%found that $B(\sg \to c t \nt_1)$ is insensitive to $\delta^{uLL}_{23}$ and can 
%be quite large ($\sim$ 50\%) in a sizable allowed range $0.03 \lsim \delta^{uLL}_{23} 
%\lsim 0.12$ as can be seen in Fig.3. This insensitivity can be explained as follows:
%the $\delta^{uLL}_{23}$ affects mainly the masses and mixing of the heavier up-type 
%squarks ${\ti u}_4$ and ${\ti u}_5$, but mainly the on-shell ${\ti u}_{1,2}$ mediate 
%the decay $\sg \to c t \nt_1$ in this scenario. 

%-----------------------------------------------------------------------------
\section{Impact on collider signatures}
%-----------------------------------------------------------------------------
%
%-----------------------------------------------------------------------------
%\section{Invariant mass distributions}
%-----------------------------------------------------------------------------

Here we study the invariant mass distributions (i.e. the differential decay branching 
ratios) ${\rm dBr}(\tilde g\to \tilde u_i u_j \to u_j u_k \tilde\chi^0_n)/{\rm d}M_{u_j u_k}$, 
with $M_{u_j u_k}$ being the invariant mass of the two quark system $u_j u_k$ in the 
final state.
The kinematical endpoinds of the distributions are given in terms of the masses 
of the involved particles by \cite{Hisano:2002xq}
\begin{eqnarray}
M^{i({\rm min,max})}_{u_j u_k}&=&\biggl\{ m^2_{u_j}+m^2_{u_k}+
\frac{1}{2 m^2_{\tilde u_i}}\left[
(m^2_{\tilde g}-m^2_{u_j}-m^2_{\tilde u_i})
(m^2_{\tilde u_i}+m^2_{u_k}-m^2_{\tilde\chi^0_n})
\right.\nonumber\\[2mm]
&&{}\left. \mp\lambda^{\frac{1}{2}}(m^2_{\tilde g},m^2_{u_j},
m^2_{\tilde u_i})~
\lambda^{\frac{1}{2}}(m^2_{\tilde u_i},m^2_{u_k},m^2_{\tilde\chi^0_n})
\right]\biggr\}^{\frac{1}{2}}~,
\label{eq:InvMass}
\end{eqnarray}
with $\lambda(x,y,z)=x^2+y^2+z^2-2(xy+xz+yz)$, 
where $\ti{u}_i$ is the intermediate squark, $u_j$ is from the primary decay 
(i.e. the two-body $\ti g$ decay) and $u_k$ is from the secondary decay 
(i.e. the $\ti{u}_i$ decay).
Note that $M^{i({\rm min,max})}_{u_j u_k} \neq M^{i({\rm min,max})}_{u_k u_j}$ 
for $j\neq k$. 
%Note that the endpoints depend on the mass of the intermediate squark ${\ti u}_i$ and 
%that $M^{i({\rm min,max})}_{u_j u_k} \neq M^{i({\rm min,max})}_{u_k u_j}$ for $j\neq k$. 
%
We calculate the invariant mass distributions by summing over the
intermediate up-type squarks giving rise to the same final state: 
\begin{eqnarray}
{\rm dBr}(\tilde g \to u_j u_k \tilde\chi^0_n)/{\rm d}M_{u_j u_k} &=& 
\frac{1}{1+\delta_{jk}} \sum_i \left[{\rm dBr}(\tilde g\to \tilde u_i u_j \to u_j u_k \tilde\chi^0_n)/{\rm d}M_{u_j u_k} 
\right. \nonumber\\[2mm]
&&{}
\left. + {\rm dBr}(\tilde g\to \tilde u_i u_k \to u_k u_j \tilde\chi^0_n)/{\rm d}M_{u_j u_k} \right]. 
\label{eq:Distri}
\end{eqnarray}
%
%We calculate the invariant mass distributions by summing over the
%intermediate up-type squarks that give rise to the same final state. 
%Then the branching ratio is given by 
%
%\begin{eqnarray}
%{\rm B}(\tilde g\to u_j u_k \tilde\chi^0_n)&=&
%\sum_i \left[\int^{M_{u_j u_k}^{i({\rm max})}}_{M_{u_j u_k}^{i({\rm min})}}
%\frac{{\rm d}{\rm Br}(\tilde g \to \tilde u_i u_j \to u_j u_k \tilde\chi^0_n)}{{\rm d}M_{u_j u_k}} 
%{{\rm d}M_{u_j u_k}}
%\right. \nonumber\\[2mm]
%&&{}
%\left.+\int^{M_{u_k u_j}^{i({\rm max})}}_{M_{u_k u_j}^{i({\rm min})}}
%\frac{{\rm d}{\rm Br}(\tilde g \to \tilde u_i u_k \to u_k u_j \tilde\chi^0_n)}{{\rm d}M_{u_j u_k}} 
%{\rm d}M_{u_j u_k}
%\right]. 
%\label{eq:Distri}
%\end{eqnarray}
% 
%The formula in Eq.~(\ref{eq:Distri}) holds for the case 
%when the quarks in the final state have different flavor, i.e. $j\neq k$.
%%The two terms in the brackets take into account that the quarks ($u_j$ and $u_k$)
%%can originate from the first or second decay. Note that the
%%matrix elements as well as the kinematical endpoinds, Eq.~(\ref{eq:Distri}), are 
%%different for the two cases.
%The formula can also be used for the case of same flavour quarks, i.e. $j=k$, 
%but then one has to divide the right-hand 
%side of Eq.~(\ref{eq:Distri}) by a factor 2.
Note that the individual distribution 
${\rm d}{\rm Br}(\tilde g \to \tilde u_i u_j \to u_j u_k \tilde\chi^0_n)/{\rm d}M_{u_j u_k}$ 
(${\rm d}{\rm Br}(\tilde g \to \tilde u_i u_k \to u_k u_j \tilde\chi^0_n)/{\rm d}M_{u_j u_k}$), 
is proportional to $M_{u_j u_k}$ and its allowed range is given by 
$[M_{u_j u_k}^{i({\rm min})},M_{u_j u_k}^{i({\rm max})}]$
($[M_{u_k u_j}^{i({\rm min})},M_{u_k u_j}^{i({\rm max})}]$). 

In the following we show how QFV due to the 2nd and 3rd generation mixing of 
the up-type squarks influences the invariant mass distributions. 
We discuss two scenarios, one with gluino mass $m_{\ti g} = 800 \gev$ and 
the other with $m_{\ti g} = 1300 \gev$. 

%%-----------------------------------------------------------------------------
%\subsection{Light gluino scenario: $m_{\tilde g}=800$~GeV}
%%-----------------------------------------------------------------------------

%We start from the $m_{\ti g} = 800 \gev$  scenario given in Table \ref{tab1}. 
We start from the QFV scenario with $m_{\ti g} = 800 \gev$ given 
in Table \ref{tab1}. In this QFV scenario the squark mass eigenstates $\tilde u_1$ and 
$\tilde u_2$ are a strong mixture of the flavour eigenstates $\tilde c_R$ and 
$\tilde t_R$. 
First we consider the invariant mass distribution for a final state 
including two top quarks. 
Fig.6 shows the invariant mass distributions of the top quark pairs for the 
QFV scenario, where one has B($\tilde g\to t \bar{t} \tilde\chi^0_1)=12.0\%$. 
Note that the invariant mass distribution of the two top quarks in the QFV scenario 
shows no additional edge structure. 
%shows no additional edge structure as compared to that in the QFC scenario. 
This is because only the lightest up-type squark, $\ti u_1$, 
can mediate this final state while the other squarks are too heavy. \\
Next we consider the invariant mass distribution for a final state including 
c and t quarks in the QFV scenario of Table \ref{tab1}, where one has 
B($\tilde g\to c t \tilde\chi^0_1)=46.3\%$. Fig.6 shows the invariant mass 
distribution of $ct$. There are more edge structures due to the processes 
$\tilde g\to \tilde u_1 t \to t c \tilde\chi^0_1$
[with $M^{1\rm(min,max)}_{tc}=(253,526)$~GeV],
$\tilde g\to \tilde u_1 c \to c t \tilde\chi^0_1$ 
[with $M^{1\rm(min,max)}_{ct}=(254,580)$~GeV], and 
$\tilde g\to \tilde u_2 c \to c t \tilde\chi^0_1$
[with $M^{2\rm(min,max)}_{ct}=(219,497)$~GeV]. 
Note that $\tilde g\to \tilde u_2 t$ is kinematically forbidden in this scenario. 
%$\tilde g\to \tilde u_1 t$; $\tilde u_1 \to \tilde\chi^0_1 c$
%[with $M^{1\rm(min,max)}_{tc}=(249?,532?)$~GeV],
%$\tilde g\to \tilde u_1 c$; $\tilde u_1 \to \tilde\chi^0_1 t$ 
%[with $M^{1\rm(min,max)}_{ct}=(248?,585?)$~GeV],
%$\tilde g\to \tilde u_2 c$; $\tilde u_2 \to \tilde\chi^0_1 t$
%[with $M^{2\rm(min,max)}_{ct}=(215?,500?)$~GeV]. 
We see that the three remarkable endpoint-edges are fairly well separated.
%We see that the three remarkable endpoint-edges around 550 GeV are fairly well separated.

%%-----------------------------------------------------------------------------
%\subsection{Scenario with $m_{\tilde g}=1300$~GeV}
%%-----------------------------------------------------------------------------

Next we consider the invariant mass distribution of final state quarks for a QFV scenario 
with a heavier gluino ($m_{\ti g}=1300 \gev$) given in Table \ref{tab4}. This scenario 
is inspired by the mSUGRA scenario A of Ref. \cite{Battaglia:2001zp} and satisfies all of 
the conditions (i)-(v) in section \ref{sec:constraints}.
%The observables of the B-physics sector have
%the following values: B($b \to s~\gamma)=3.76\times 10^{-4}$, 
%B($b \to s ~ l^+l^-)=1.60\times 10^{-6}$, B($B_s \to \mu^+\mu^-)=4.76\times 10^{-9}$, 
%$B(B^+_u \to \tau^+ \nu) = 7.90 \times 10^{-5}$, $\Delta M_{B_s}=17.25$ ps$^{-1}$.
The resulting masses of squarks, neutralinos and charginos  are given in Table \ref{tab5}.
We show the corresponding up-type squark compositions in the flavour eigenstates 
in Table \ref{tab6}. 
In this scenario the squark mass eigenstate $\tilde u_1$ ($\tilde u_2$) 
is dominated by a strong mixture of the flavour eigenstates $\tilde t_R$ and $\tilde c_R$ 
($\tilde t_L$ and $\tilde c_L$ ). 
In Fig.7 we show the two invariant mass distributions of $tt$ and $ct$, 
where one has B($\tilde g\to t t \tilde\chi^0_1)=16.6\%$,
and B($\tilde g\to c t \tilde\chi^0_1)=31.4\%$. 
Note that the QFV decay branching ratio B($\tilde g\to c t \tilde\chi^0_1)$ is large. \\
%The invariant mass distribution of two top quarks shows no additional edge structure 
%as in the scenario with $m_{\ti g}=800 \gev$ above. 
%This is because practically only the lightest up-type squark, $\ti u_1$, can mediate 
%this final state while the other up-type squarks are too heavy. 
The invariant mass distribution of two top quarks shows no additional edge structure 
for the same reason as in the scenario with $m_{\ti g}=800 \gev$ discussed above.
The decay $\tilde g\to \tilde u_2 t$ is kinematically allowed but phase-space suppressed. 
Moreover, $\ti u_2 \to t \nt_1$ is strongly suppressed because $\ti u_2 (\sim \ti t_L + \ti c_L)$ 
does not significantly couple to $\nt_1(\sim \ti B^0 (\rm Bino))$ in this scenario. 
Hence, B($\tilde g \to \tilde u_2 t \to t t \nt_1$)(=0.00035) is very small.\\
As for the invariant mass distribution of c and t quarks in the QFV scenario of Table 
\ref{tab4}, there are more edge structures due to the $\ti u_1$-mediated processes 
$\tilde g\to \tilde u_1 t \to t c \tilde\chi^0_1$
[with $M^{1\rm(min,max)}_{tc}=(601,971)$~GeV], and 
$\tilde g\to \tilde u_1 c \to c t \tilde\chi^0_1$ 
[with $M^{1\rm(min,max)}_{ct}=(183,1022)$~GeV]. 
The decays $\tilde g\to \tilde u_2 ~ c/t$ are phase-space suppressed and 
the decays $\ti u_2 \to ~ c/t ~ \nt_1$ are strongly suppressed in this 
scenario as is explained above. 
Hence, B($\tilde g \to \tilde u_2 ~ c/t \to c t \nt_1$)(=0.0004) is very small.
%Notice a structure around 600 GeV.
%
%Note that there is a remarkable structure around 600 GeV.
%
\begin{table}[t]
\begin{center}

\begin{tabular}{|c||c|c|c|} \hline
 $M^2_{Q\alpha\beta}$
& \multicolumn{1}{c|}{\scriptsize{${\beta=1}$}} 
& \multicolumn{1}{c|}{\scriptsize{$\beta=2$}} 
& \multicolumn{1}{c|}{\scriptsize{$\beta=3$}} \\\hline\hline
 \scriptsize{$\alpha=1$}
& \multicolumn{1}{c|}{$(1200)^2$} 
& \multicolumn{1}{c|}{0} 
& \multicolumn{1}{c|}{0} \\\hline

 \scriptsize{$\alpha=2$}
& \multicolumn{1}{c|}{0} 
& \multicolumn{1}{c|}{$(1200)^2$} 
& \multicolumn{1}{c|}{$(500)^2$} \\\hline

 \scriptsize{$\alpha=3$}
& \multicolumn{1}{c|}{0} 
& \multicolumn{1}{c|}{$(500)^2$} 
& \multicolumn{1}{c|}{$(1128)^2$} \\\hline
\end{tabular}
\hspace{0.1cm}
\begin{tabular}{|c|c|c|c|c|c|} \hline
 
  \multicolumn{1}{|c|}{$M_1$} 
& \multicolumn{1}{c|}{$M_2$} 
& \multicolumn{1}{c|}{$m_{\sg}$} 
& \multicolumn{1}{c|}{$\mu$} 
& \multicolumn{1}{c|}{$\tan\beta$} 
& \multicolumn{1}{c|}{$m_{A^0}$} \\\hline\hline
 
  \multicolumn{1}{|c|}{255} 
& \multicolumn{1}{c|}{497} 
& \multicolumn{1}{c|}{1300} 
& \multicolumn{1}{c|}{756} 
& \multicolumn{1}{c|}{5} 
& \multicolumn{1}{c|}{800} \\\hline

\end{tabular}
\vskip0.2cm
\begin{tabular}{|c||c|c|c|} \hline
 $M^2_{D\alpha\beta}$
& \multicolumn{1}{c|}{\scriptsize{${\beta=1}$}} 
& \multicolumn{1}{c|}{\scriptsize{$\beta=2$}} 
& \multicolumn{1}{c|}{\scriptsize{$\beta=3$}} \\\hline\hline
 \scriptsize{$\alpha=1$}
& \multicolumn{1}{c|}{$(1141)^2$} 
& \multicolumn{1}{c|}{0} 
& \multicolumn{1}{c|}{0} \\\hline

 \scriptsize{$\alpha=2$}
& \multicolumn{1}{c|}{0} 
& \multicolumn{1}{c|}{$(1141)^2$} 
& \multicolumn{1}{c|}{0} \\\hline

 \scriptsize{$\alpha=3$}
& \multicolumn{1}{c|}{0} 
& \multicolumn{1}{c|}{0} 
& \multicolumn{1}{c|}{$(1100)^2$} \\\hline
\end{tabular}
\hspace{0.4cm}
\begin{tabular}{|c||c|c|c|} \hline
 $M^2_{U\alpha\beta}$
& \multicolumn{1}{c|}{\scriptsize{${\beta=1}$}} 
& \multicolumn{1}{c|}{\scriptsize{$\beta=2$}} 
& \multicolumn{1}{c|}{\scriptsize{$\beta=3$}} \\\hline\hline
 \scriptsize{$\alpha=1$}
& \multicolumn{1}{c|}{$(1149)^2$} 
& \multicolumn{1}{c|}{0} 
& \multicolumn{1}{c|}{0} \\\hline

 \scriptsize{$\alpha=2$}
& \multicolumn{1}{c|}{0} 
& \multicolumn{1}{c|}{$(1149)^2$} 
& \multicolumn{1}{c|}{$(894)^2$} \\\hline

 \scriptsize{$\alpha=3$}
& \multicolumn{1}{c|}{0} 
& \multicolumn{1}{c|}{$(894)^2$} 
& \multicolumn{1}{c|}{$(877)^2$} \\\hline
\end{tabular}
%\\[0.5ex]
\vskip0.4cm
\caption{\label{tab4}
The MSSM parameters in the QFV scenario with $m_{\ti g}=1300 \gev$. 
All of $A_{U \a\b}$ and $A_{D \a\b}$ are set to zero.
All mass parameters are given in GeV.}
\end{center}
\end{table}

\begin{table}[t]
\begin{center}

\begin{tabular}{|c|c|c|c|c|c|} \hline
 
  \multicolumn{1}{|c|}{$\tilde u_1$} 
& \multicolumn{1}{c|}{$\tilde u_2$} 
& \multicolumn{1}{c|}{$\tilde u_3$} 
& \multicolumn{1}{c|}{$\tilde u_4$} 
& \multicolumn{1}{c|}{$\tilde u_5$} 
& \multicolumn{1}{c|}{$\tilde u_6$} \\\hline\hline
 
  \multicolumn{1}{|c|}{466} 
& \multicolumn{1}{c|}{1054} 
& \multicolumn{1}{c|}{1149} 
& \multicolumn{1}{c|}{1199} 
& \multicolumn{1}{c|}{1275} 
& \multicolumn{1}{c|}{1379} \\\hline

\end{tabular}
\begin{tabular}{|c|c|c|c|c|c|} \hline
 
  \multicolumn{1}{|c|}{$\tilde d_1$} 
& \multicolumn{1}{c|}{$\tilde d_2$} 
& \multicolumn{1}{c|}{$\tilde d_3$} 
& \multicolumn{1}{c|}{$\tilde d_4$} 
& \multicolumn{1}{c|}{$\tilde d_5$} 
& \multicolumn{1}{c|}{$\tilde d_6$} \\\hline\hline
 
  \multicolumn{1}{|c|}{1046} 
& \multicolumn{1}{c|}{1101} 
& \multicolumn{1}{c|}{1141} 
& \multicolumn{1}{c|}{1141} 
& \multicolumn{1}{c|}{1201} 
& \multicolumn{1}{c|}{1274} \\\hline

\end{tabular}
\vskip0.2cm
\begin{tabular}{|c|c|c|c||c|c|} \hline
 
  \multicolumn{1}{|c|}{$\tilde \chi^0_1$} 
& \multicolumn{1}{c|}{$\tilde \chi^0_2$} 
& \multicolumn{1}{c|}{$\tilde \chi^0_3$} 
& \multicolumn{1}{c||}{$\tilde \chi^0_4$} 
& \multicolumn{1}{c|}{$\tilde \chi^\pm_1$} 
& \multicolumn{1}{c|}{$\tilde \chi^\pm_2$} \\\hline\hline
 
  \multicolumn{1}{|c|}{253} 
& \multicolumn{1}{c|}{483} 
& \multicolumn{1}{c|}{758} 
& \multicolumn{1}{c||}{775} 
& \multicolumn{1}{c|}{482} 
& \multicolumn{1}{c|}{774} \\\hline

\end{tabular}
\vskip0.4cm
\caption{\label{tab5}
Sparticles and corresponding masses (in GeV) in the scenario of 
Table \ref{tab4}.
}
\end{center}
\end{table}
\begin{table}[t]
\begin{center}
%\hspace{2mm}   
\begin{tabular}{|c||c|c|c|c|c|c|}
		 \hline
			$R^{\tilde u}_{i\a}$
		  & $\ti u_L$ & $\ti c_L$ & $\ti t_L$ & $\ti u_R$ & $\ti c_R$ & $\ti t_R$ \\
		 \hline\hline
		  $\ti u_1$  & -0.001 & 0.006 & -0.021 & 0 & 0.587 & -0.809 \\
		  $\ti u_2$  & -0.137 & 0.621 & -0.771 & 0 & -0.024 & 0.006 \\
		  $\ti u_3$  & 0 & 0 & 0 & -1.0 & 0 & 0 \\
		  $\ti u_4$  & -0.976 & -0.219 & -0.003 & 0 & 0 & 0 \\
		  $\ti u_5$  & 0.171 & -0.752 & -0.636 & 0 & -0.032 & -0.012 \\
		  $\ti u_6$  & 0.003 & -0.015 & -0.033 & 0 & 0.808 & 0.588 \\

		 \hline
		 \end{tabular} 
		 
\vspace{3mm} 
%
%\begin{tabular}{|c||c|c|c|c|c|c|}
%		 \hline
%			$R^{\tilde d}_{i\a}$
%		  & $\ti d_L$ & $\ti s_L$ & $\ti b_L$ & $\ti d_R$ & $\ti s_R$ & $\ti b_R$ \\
%		 \hline\hline
%		  $\ti d_1$  & 0 & -0.582 & 0.810 & 0 & -0.001 & 0.076 \\
%		  $\ti d_2$  & 0 & -0.057 & 0.053 & 0 & 0 & -0.997 \\
%		  $\ti d_3$  & 0 & 0 & 0 & 0 & 1.0 & 0 \\
%		  $\ti d_4$  & 0 & 0 & 0 & 1.0 & 0 & 0 \\
%		  $\ti d_5$  & -1.0 & 0 & 0 & 0 & 0 & 0 \\
%		  $\ti d_6$  & 0 & -0.811 & -0.584 & 0 & 0.001 & 0.015 \\
%
%		 \hline
%		 \end{tabular} 

	\caption{\label{tab6} The up-type squark compositions in the flavour eigenstates, 
i.e. the mixing matrix $R^{\ti u}_{i\a}$ for the scenario of Table \ref{tab4}.}

\end{center}
\end{table}

Finally, we briefly discuss the measurability of the QFV decay
$\sg \to$ c t $\nt_1$ at LHC. 
It is important whether one can discriminate between the QFV 
decay $\sg \to$ c t $\nt_1$ and the QFC decay $\sg \to$ t t $\nt_1$. 
Therefore, it is necessary to identify the top quarks 
in the final states. This is possible by using the decay $t \to b W$ 
with the W decaying into two jets. For this purpose, a special method 
was proposed in \cite{Hisano:2002xq}, where it is assumed that the masses 
of the gluino and the $\nt_1$ are known from other measurements. 
The signature of the decay $\sg \to$ c t $\nt_1$ would be
'charm-jet + top-quark + missing-energy'. Therefore, charm-tagging also 
would be very useful. If this is not possible, one should search for the
decay $\sg \to$ q t $\nt_1$ (q $\neq$ t), i.e. for the signature 'jet +  
top-quark + missing-energy'. In the scenarios discussed, the most important
SUSY background would be due to the QFC decay $\sg \to$ t t $\nt_1$ and 
the pair production of the lightest up-type squarks,
$p p \to {\ti u}_1 + {\ti u}_1 + X$, with ${\ti u}_1 \to c \nt_1$ and 
${\ti u}_1 \to t \nt_1$. 
The most important SM background would be top-quark pair production. 
For the measurement of the endpoints in the multiple edge structure a good energy/momentum
resolution of the detector would be necessary.
In any case, one should take into account the possibility of significant
contributions from QFV decays in the gluino search. 
Moreover one should also include the QFV squark parameters in the determination of 
the basic SUSY parameters at LHC. It is clear that detailed Monte Carlo studies 
taking into account backgrounds and detector simulations would be necessary. 
Such studies are beyond the scope of the present article.

%-------------------------------------------------------------------
\section{Conclusion \label{Conclusion}}
%-------------------------------------------------------------------

To conclude, we have studied gluino decays in the MSSM with squark mixing of the 
second and third generation, especially $\ti c_{L/R}$ - $\ti t_{L/R}$ mixing. 
We have shown that QFV gluino decay branching ratios such as B($\sg \to$ c t $\nt_1$) 
can be very large due to the squark mixing in a significant part of the MSSM 
parameter space despite the very strong experimental constraints from B factories, Tevatron 
and LEP with those of $b \to s \gamma$ and $\Delta M_{B_s}$ being especially important.\\
We have also studied the effect of the squark generation mixing on the invariant mass 
distributions of the two quarks from the gluino decay at LHC. We have found that 
it can result in novel and characteristic edge structures in the distributions. 
In particular, multiple-edge (3- or 4-edge) structures can appear in the 
charm-top quark mass distribution. 
The appearance of these remarkable structures would 
provide an additional powerful test of supersymmetric QFV at LHC.\\ 
These could have an important impact on the search for gluinos and the MSSM parameter 
determination at LHC.

\section*{Acknowledgments}

This work is supported by the "Fonds zur F\"orderung der
wissenschaftlichen Forschung (FWF) of Austria, project No. P18959-N16.
The authors acknowledge support from EU under the MRTN-CT-2006-035505
and MRTN-CT-2006-503369 network programs. 
%and MTRN-CT-2006-503369 network proprammes. 
A. B. was supported by the
Spanish grants SAB 2006-0072, FPA 2005-01269, FPA 2005-25348-E and FPA
2008-00319/FPA of the Ministero de Educacion y Ciencia.
T. K. is supported by the Portuguese FCT through the projects
POCI/FP/81919/2007 and CFTP-FCT UNIT 777, which are partially
funded through POCTI (FEDER). 
W. P. is partially supported by the German Ministry of Education and Research 
(BMBF) project nr. 05HT6WWA.

%--------------------------------

%--------------------------------

\newpage

%------------------------------------------------------------------------
% Figures Captions
%------------------------------------------------------------------------
\begin{flushleft}
{\Large \bf Figure Captions} \\
\end{flushleft}

\noi
{\bf Figure 1}: 
Feynman diagrams for $\tilde g \to \tilde u_i~c \to c~t~ \tilde\chi^0_{1}$ (left) 
and $\tilde g \to \tilde u_i~t \to c~t~ \tilde\chi^0_{1}$ (right).

\noi
{\bf Figure 2}: 
Contours of the QFV decay branching ratio $B(\sg \to c t \nt_1)$ in the 
($\Delta M^2_U, M^2_{U 23}$) plane where all of the conditions 
(i)-(v) are satisfied. 
%All basic parameters other than $M^2_{U 22}$ and $M^2_{U 23}, 
%M^2_{U 32}$ are fixed as in our reference scenario defined in Table \ref{tab1}.
The point "x" of $(\Delta M^2_U, M^2_{U 23}) = (2.36 \times 10^4, 5 \times 10^4) 
\gev^2$ corresponds to our reference scenario of Table \ref{tab1}.

\noi 
{\bf Figure 3}:
Contours of the QFV decay branching ratio $B(\sg \to c t \nt_1)$ (solid lines) in the 
$\delta^{uLL}_{23} - \delta^{uRR}_{23}$ plane where all of the conditions (i)-(v) 
except the $b \to s \gamma$ constraint are satisfied. 
Contours of $10^4 \times B(b \to s \gamma)$ (dashed lines) are also shown. The condition (i) 
requires $3.03 <  10^4 \times B(b \to s ~\gamma) < 4.01$. 
%All basic parameters other than $M^2_{Q 23}, M^2_{Q 32}$ and $M^2_{U 23}, M^2_{U 32}$ 
%are fixed as in our reference scenario of Table \ref{tab1}. 
The point "x" of $(\delta^{uLL}_{23}, \delta^{uRR}_{23}) = (0.068, 0.144)$ corresponds 
to our reference scenario of Table \ref{tab1}.

\noi 
{\bf Figure 4}: 
$\delta^{uRR}_{23}$ dependences of the branching ratios of (a) the gluino cascade decays, 
(b) the gluino two-body decays and (c) the up-type squark two-body decays. 
%(b) the gluino two-body decays and (c) the up-type squark two-body decays, where all 
%basic parameters other than $M^2_{U 23}, M^2_{U 32}$ are fixed as in our reference 
%scenario specified in Table \ref{tab1}.
The point "x" of $\delta^{uRR}_{23} = 0.144$ 
corresponds to our reference scenario of Table \ref{tab1}. 
The shown range of $\delta^{uRR}_{23}$ is the whole 
range allowed by the conditions (i) to (v) given in the text; note that the range 
$|\delta^{uRR}_{23}| \gsim 1.0$ is excluded by the condition $m_{\ti{u_1}} > m_{\nt_1}$ 
in (iii).

\noi 
{\bf Figure 5}: 
$\delta^{uRL}_{23}$ dependences of the branching ratios of the gluino cascade decays. 
%decays, where all basic parameters other than $A_{U 32}$ are fixed as in our reference scenario 
%specified in Table \ref{tab1}. 
The point "x" of $\delta^{uRL}_{23} = 0$ corresponds 
to our reference scenario of Table \ref{tab1}. The shown range of $\delta^{uRL}_{23}$ 
is the whole range allowed by the conditions (i) to (v) given in the text; note that the 
range $|\delta^{uRL}_{23}| \gsim 0.3$ is excluded by the condition (v).

\noi 
{\bf Figure 6}: 
Invariant mass distributions of two up-type quarks from the decay 
$\tilde g\to u_j u_k \tilde\chi^0_1$ for the QFV scenario of Table \ref{tab1}. 

\noi 
{\bf Figure 7}: 
Invariant mass distributions of two up-type quarks from the decay 
$\tilde g\to u_j u_k \tilde\chi^0_1$ for the QFV scenario of Table \ref{tab4}. 

\newpage
%
%------------------------------------------------------------------------
% Figures
%------------------------------------------------------------------------
%
% Figure 1 --------------------------------------------------------------
%
\begin{figure}[!htb] 
\begin{center}
\scalebox{1.0}[1.0]{\includegraphics{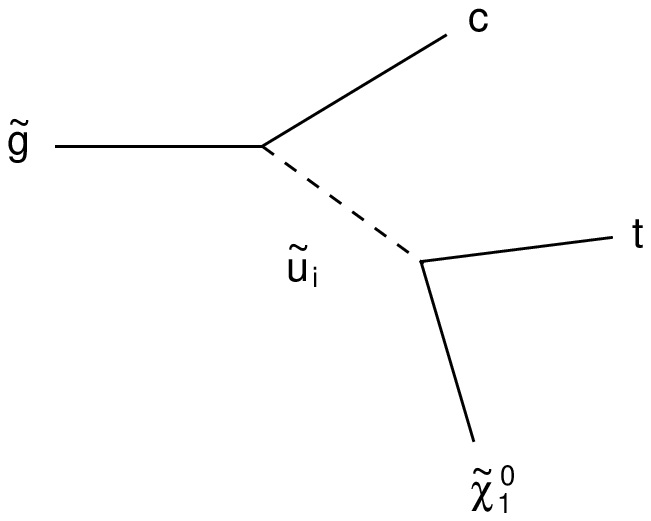}} \hspace{5mm}
\scalebox{1.0}[1.0]{\includegraphics{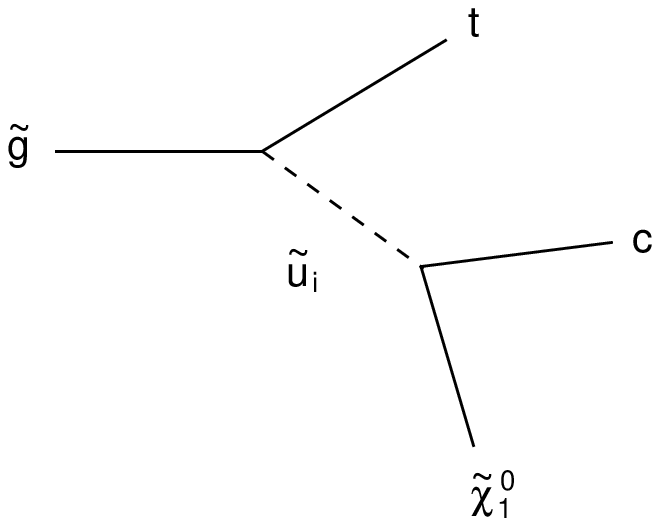}} \\
\vspace{5mm}
{\LARGE \bf Fig.1}
\end{center}
\end{figure}

\newpage
%
% Figure 2 --------------------------------------------------------------
%
\begin{figure}[!htb] 
\begin{center}
\scalebox{1.0}[1.2]{\includegraphics{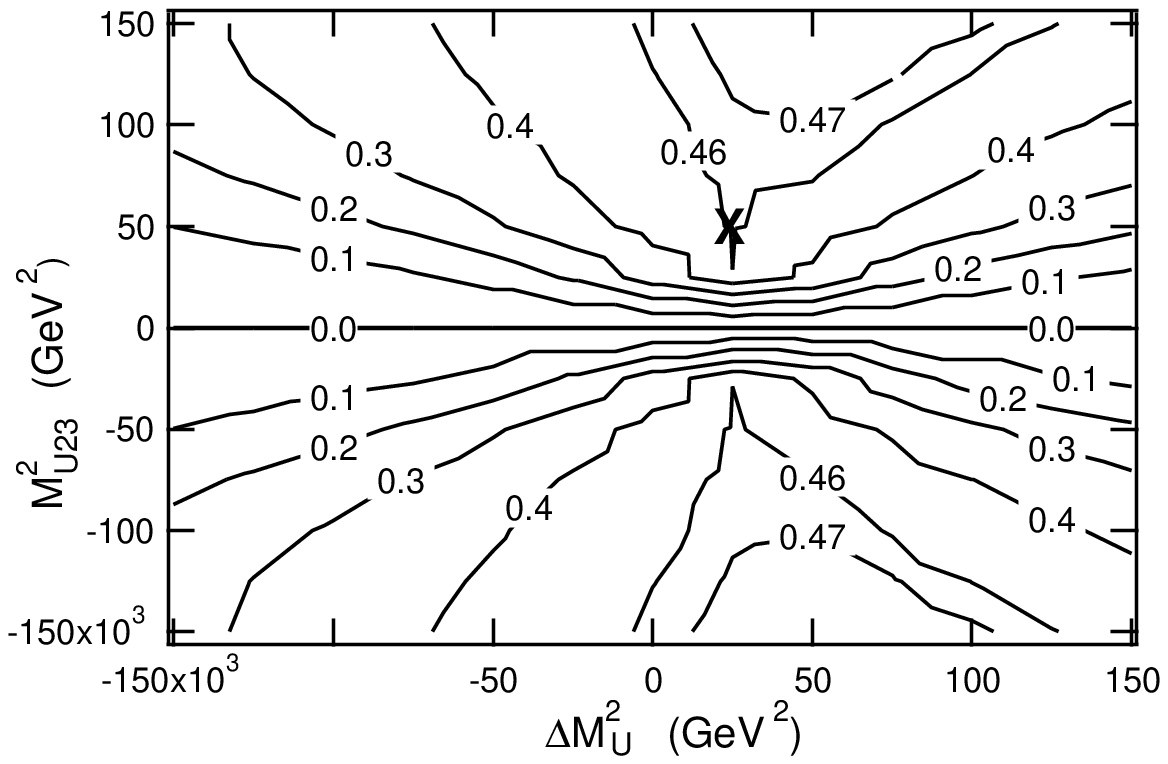}} \\
%\scalebox{0.7}[0.8]{\includegraphics{fig.2.eps}} \\
%
%\vspace{5mm}
{\LARGE \bf Fig.2}
\end{center}
\end{figure}

%\newpage
%
% Figure 3 ------------------------------------------------------------
%
\begin{figure}[!htb] 
\begin{center}
\scalebox{0.9}[0.9]{\includegraphics{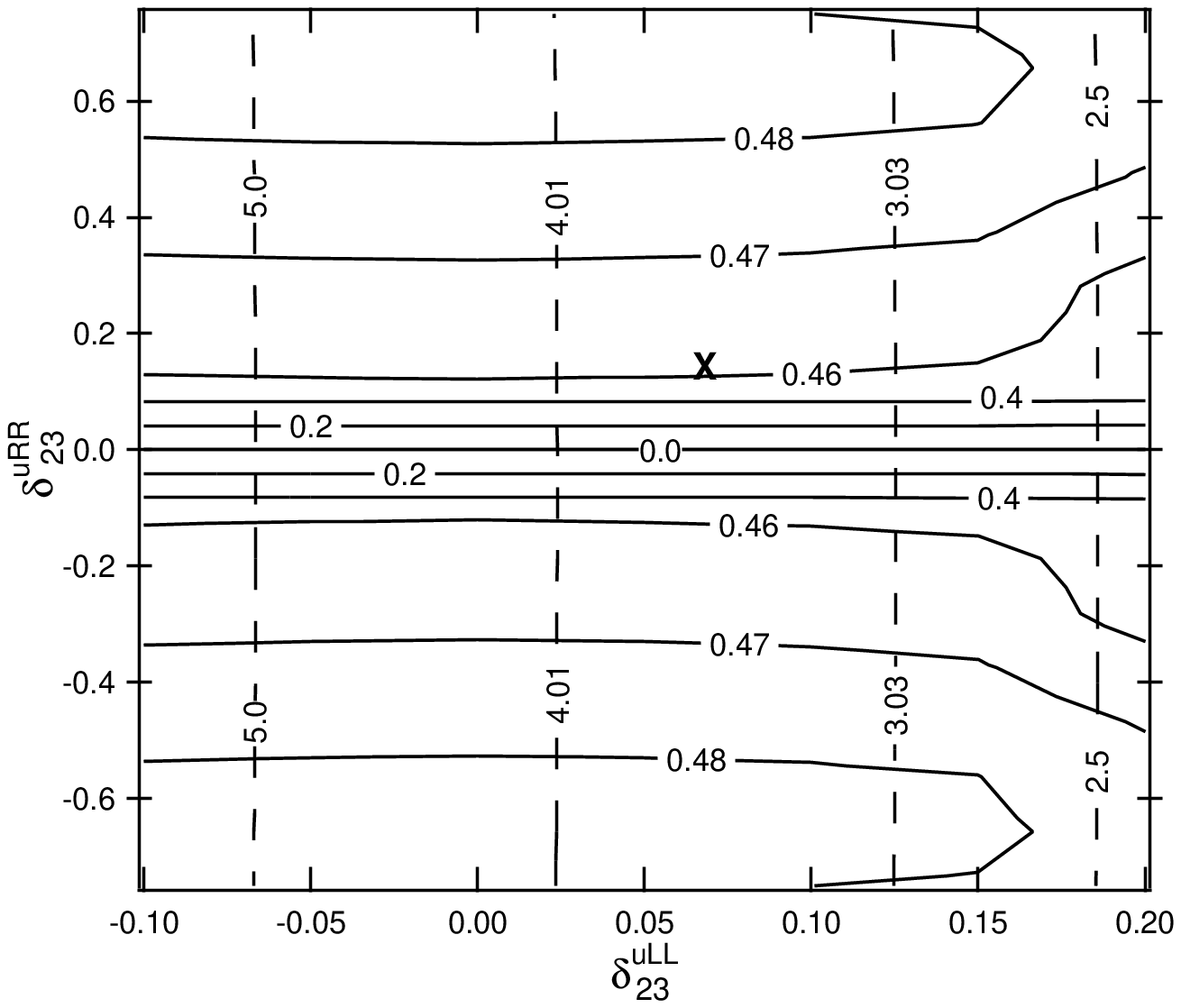}} \\ 
%\scalebox{0.7}[0.9]{\includegraphics{fig.3.eps}} \\ 
%\vspace{-5mm}
\end{center}
\end{figure}
\vspace{-10mm}
\begin{center}
{\LARGE \bf Fig.3}
\end{center}

\newpage
%
% Figure 4 ------------------------------------------------------------
%
\begin{figure}[!htb] 
\begin{center}
\scalebox{0.65}[0.85]{\includegraphics{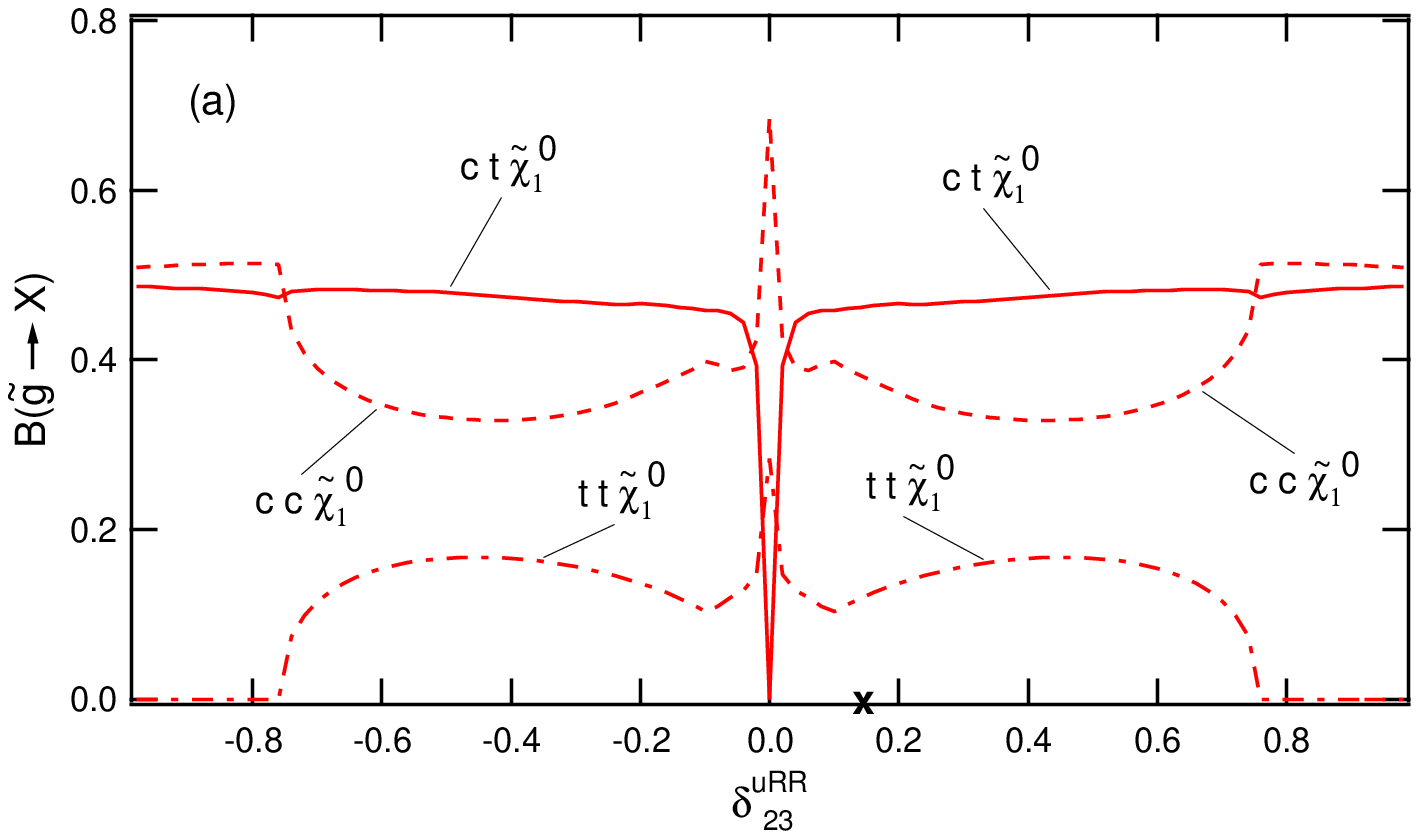}} \\
%\vspace{3mm}
\scalebox{0.65}[0.85]{\includegraphics{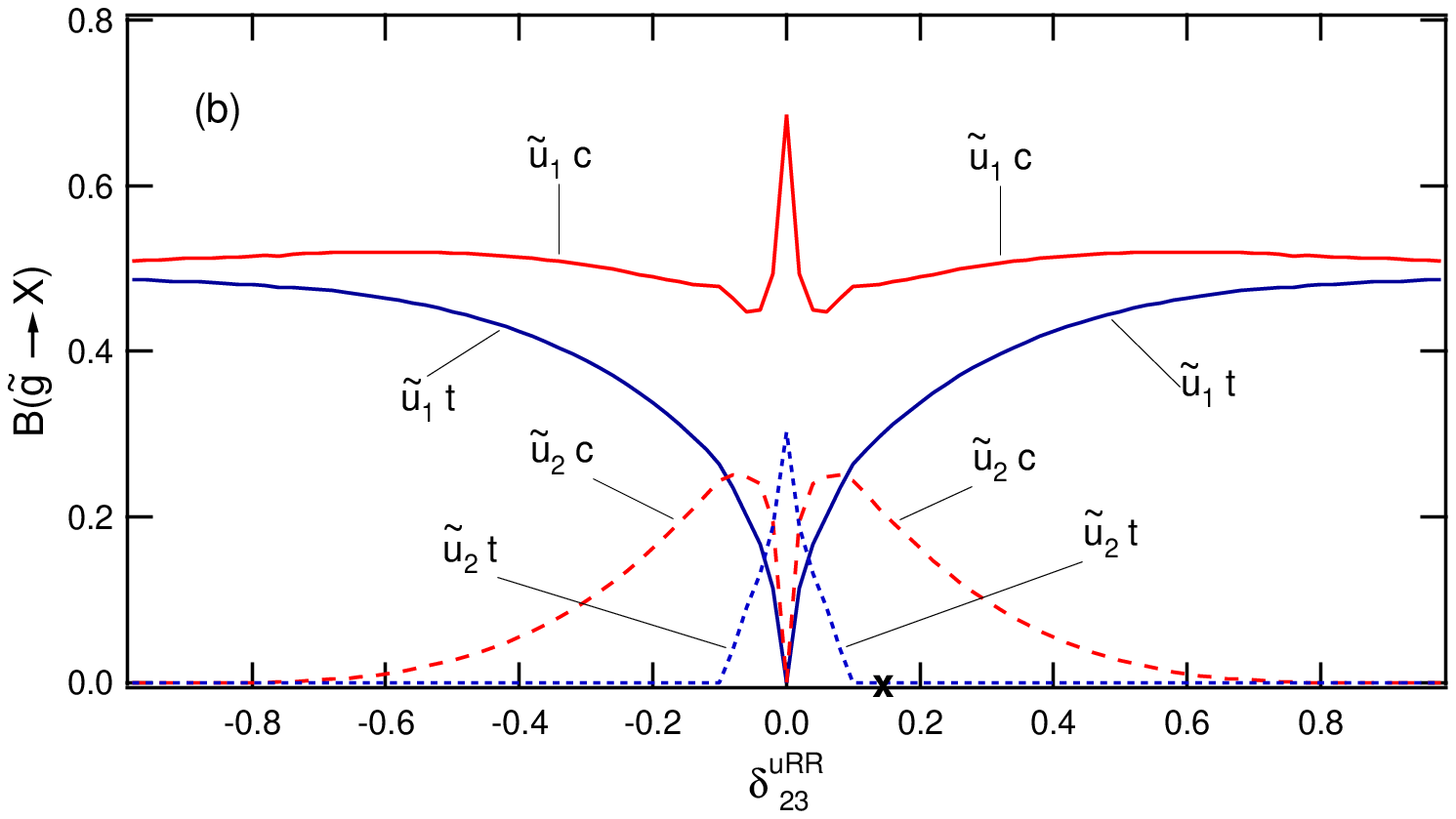}} \\
%\vspace{3mm}
\scalebox{0.65}[0.85]{\includegraphics{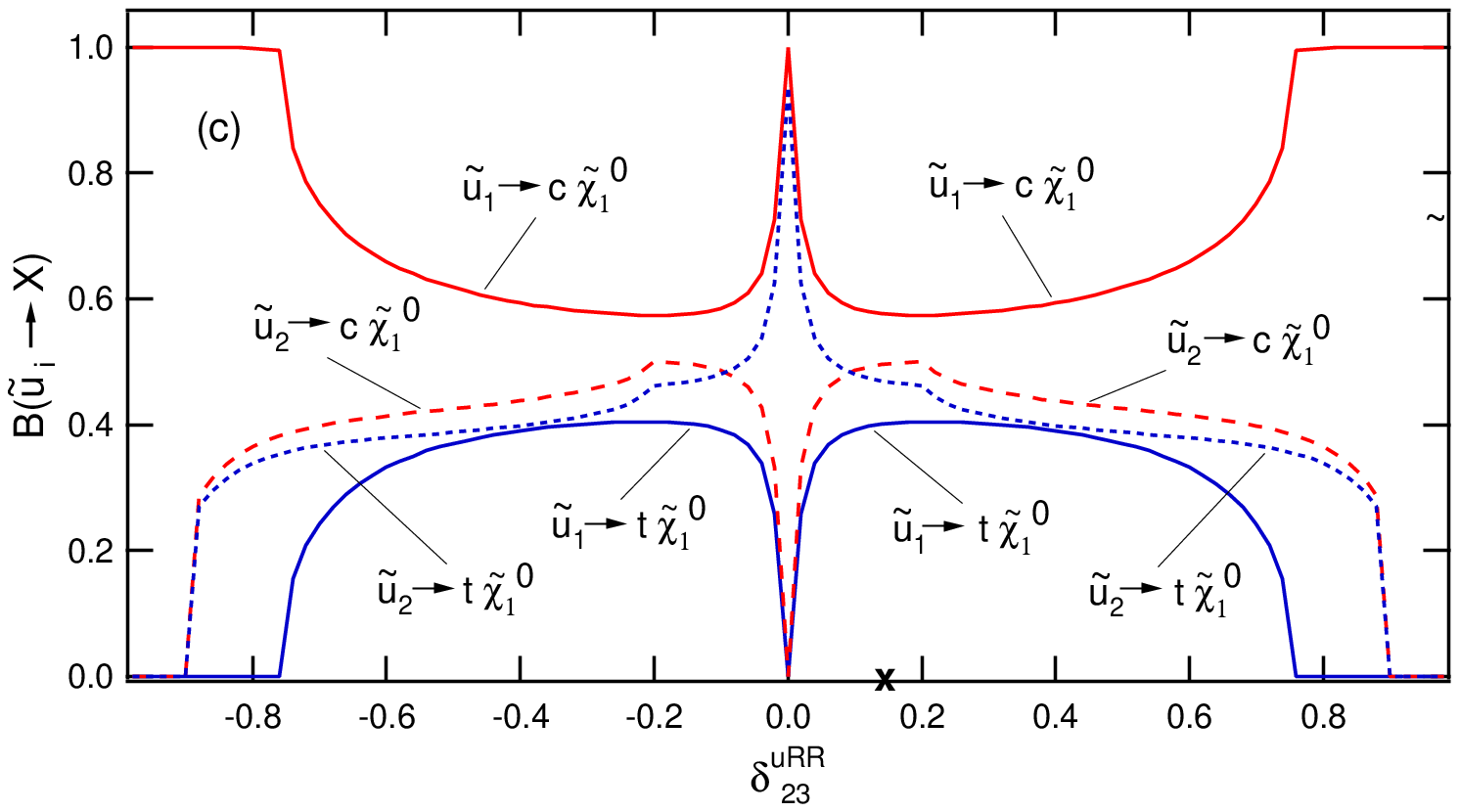}} \\
\vspace{-10mm}
\end{center}
\end{figure}

\begin{center}
{\LARGE \bf Fig.4}
\end{center}

\newpage
%
% Figure 5 ------------------------------------------------------------
%
\begin{figure}[!htb] 
\begin{center}
\scalebox{1.0}[1.0]{\includegraphics{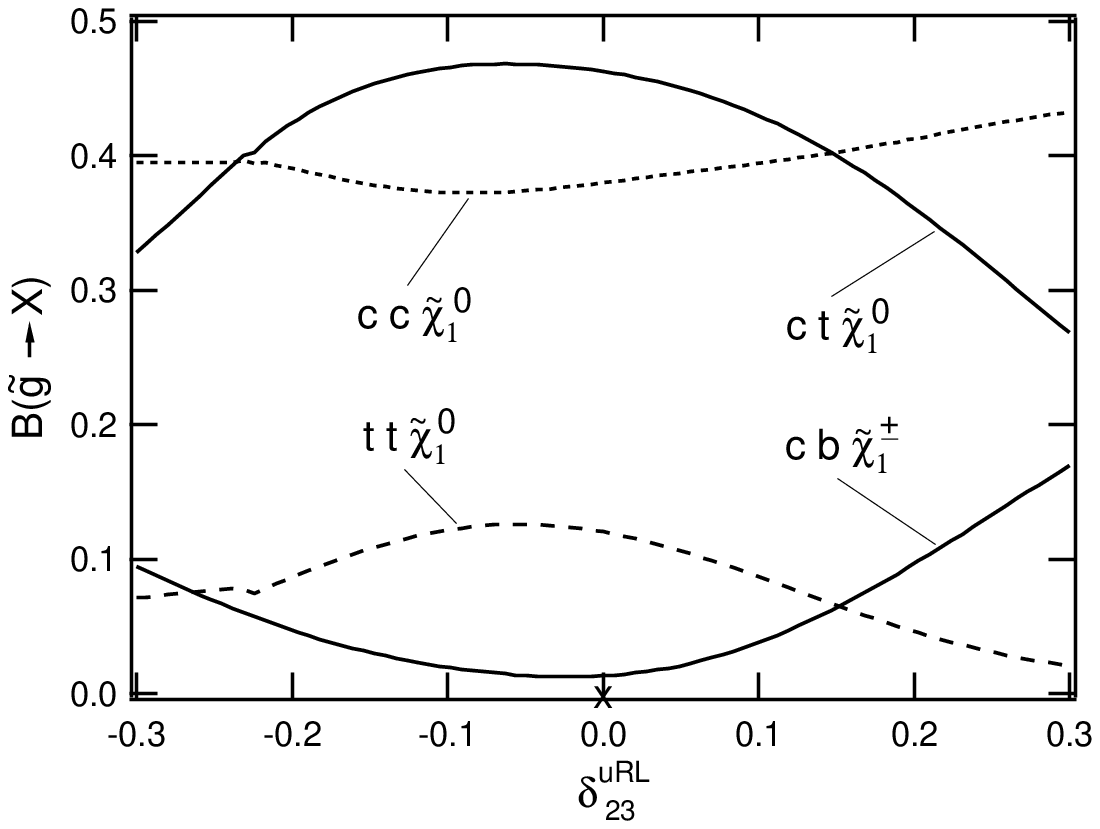}} \\
%\vspace{5mm}
{\LARGE \bf Fig.5}
\end{center}
\end{figure}

\newpage
%
% Figure 6 ------------------------------------------------------------
%
\begin{figure}[!htb] 
\begin{center}
\scalebox{1.1}[1.1]{\includegraphics{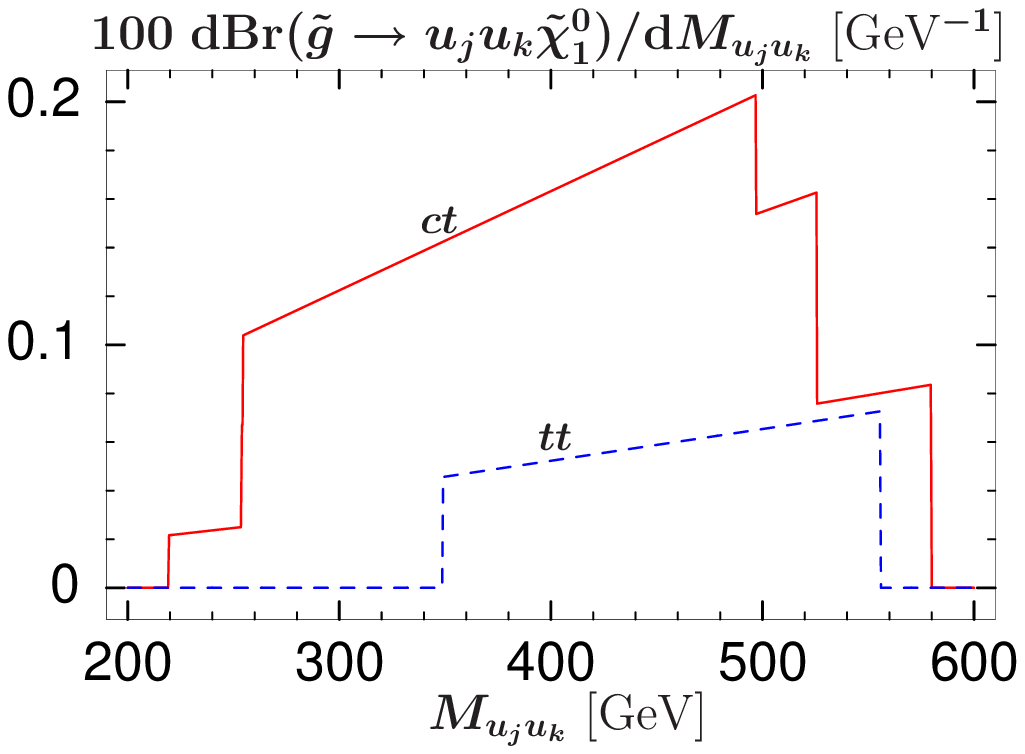}} \\ 
\vspace{5mm}
{\LARGE \bf Fig.6}
\end{center}
\end{figure}
%
%\newpage
%
% Figure 7 ------------------------------------------------------------
%
\begin{figure}[!htb] 
\begin{center}
%\hspace{25mm}
\scalebox{1.1}[1.2]{\includegraphics{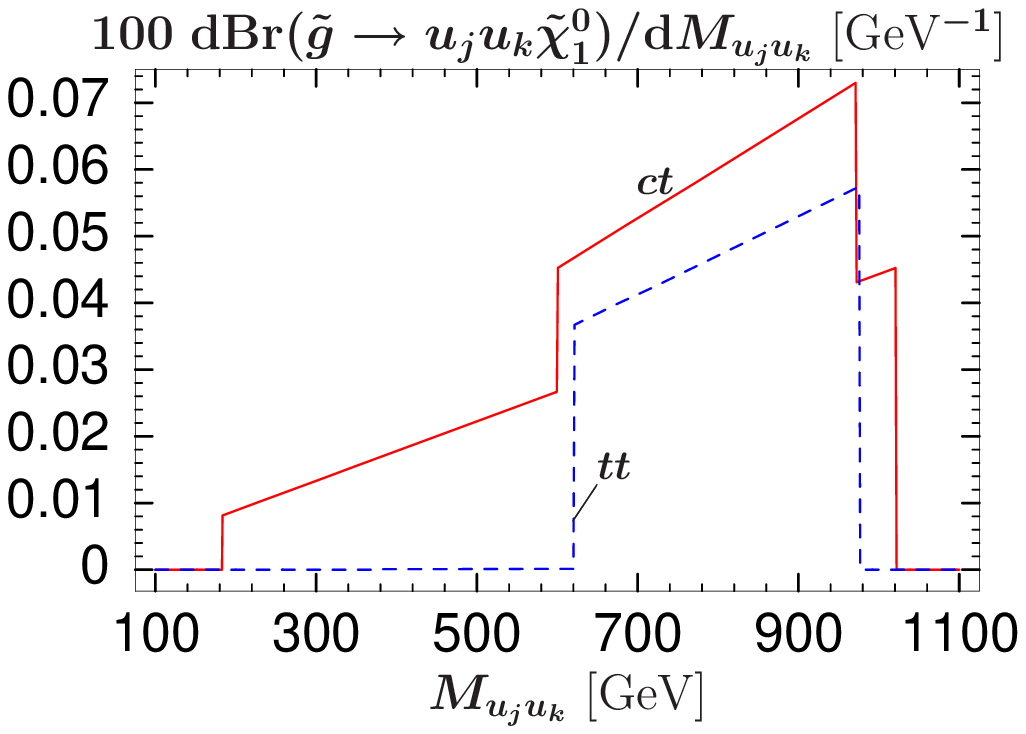}} \\ 
\vspace{5mm}
{\LARGE \bf Fig.7}
\end{center}
\end{figure}

\end{document}